\documentclass[fleqn,usenatbib]{mnras}

\usepackage{newtxtext,newtxmath}

\usepackage[T1]{fontenc}

\DeclareRobustCommand{\VAN}[3]{#2}
\let\VANthebibliography\thebibliography
\def\thebibliography{\DeclareRobustCommand{\VAN}[3]{##3}\VANthebibliography}

\usepackage{xcolor}
\usepackage{multirow}
\usepackage{soul}
\usepackage{subcaption}
\usepackage{siunitx}
\usepackage{orcidlink}
\usepackage{scrextend}

\usepackage{graphicx}	
\usepackage{amsmath}	

\defcitealias{petersson2023}{P23}
\defcitealias{zheng2020}{Z20}

\title[Resolved star-formation histories of PSB galaxies]{Spatially resolved star-formation histories of local post-starburst galaxies: Starburst and quenching spatial patterns consistent with recent mergers}

\author[H-H. Leung et al.]{
Ho-Hin Leung\orcidlink{0000-0003-0486-5178}$^{1,2}$\thanks{E-mail: hleung2@roe.ac.uk},
Vivienne Wild\orcidlink{0000-0002-8956-7024}$^{1}$,
Michail Papathomas\orcidlink{0000-0002-5897-695X}$^{3}$,
Daniel J. Mortlock\orcidlink{0000-0002-0041-3783}$^{4,5}$, \and \
Amy L. Rankine\orcidlink{0000-0002-2091-1966}$^{2}$,
Emma Curtis-Lake\orcidlink{0000-0002-9551-0534}$^{6}$, 
Yirui Zheng\orcidlink{0000-0001-7707-5930}$^{7,8}$,
Adam C. Carnall\orcidlink{0000-0002-1482-5818}$^{2}$, and \and \
Peter H. Johansson\orcidlink{0000-0001-8741-8263}$^{9}$
\\
$^{1}$SUPA\footnote{Scottish Universities Physics Alliance}, School of Physics \& Astronomy, 
University of St Andrews, 
North Haugh, St Andrews, Fife KY16 9SS, UK\\
$^{2}$SUPA, Institute for Astronomy, University of Edinburgh, Royal Observatory, Edinburgh EH9 3HJ, UK\\
$^{3}$School of Mathematics and Statistics, University of St Andrews, 
North Haugh, St Andrews, 
Fife, KY16 9SS, UK\\
$^{4}$Department of Physics, Imperial College London, Blackett Laboratory, Prince Consort Road, London SW7 2AZ, UK\\
$^{5}$Department of Mathematics, Imperial College London, London, SW7 2AZ, UK\\
$^{6}$Centre for Astrophysics Research, Department of Physics, Astronomy and Mathematics, University of Hertfordshire, Hatfield AL10 9AB, UK\\
$^{7}$School of Physics and Astronomy, Anqing Normal University, Anqing, People’s Republic of China\\
$^{8}$Institute of Astronomy and Astrophysics, Anqing Normal University, Anqing, People’s Republic of China\\
$^{9}$Department of Physics, University of Helsinki, Gustaf H\"allstr\"omin katu 2, FI-00014 Helsinki, Finland\\
}

\date{Accepted XXX. Received YYY; in original form ZZZ}

\pubyear{\the\year{}}

\begin{document}
\label{firstpage}
\pagerange{\pageref{firstpage}--\pageref{lastpage}}
\maketitle

\begin{abstract}
Post-starburst (PSB) galaxies, having recently experienced a starburst followed by rapid quenching, are excellent laboratories to probe physical mechanisms that drive starbursts and shutting down of star formation. 
Integral-field spectroscopy reveals the galaxies' spatially-resolved properties, where observed directional patterns can be linked to the galaxies' past evolution.
We measure the resolved star-formation histories (SFHs), stellar metallicity evolution and dust properties of three local PSBs from the MaNGA survey, down to $0.5${\arcsec} resolution ($\sim0.3\,$kpc) using a hierarchical Bayesian model. 
Local parameters were constrained simultaneously with parameters describing spatial trends. 
We found that all three galaxies first experienced an outer, weaker and slower quenching starburst, followed by a central, stronger and faster quenching starburst that peaked $\sim 1\,$Gyr after the first. 
The central starbursts induced a significantly stronger rise in stellar metallicity compared to the outer starbursts. 
These results are consistent with the effects of a recent gas-rich (wet) merger, where the first pericentre passage triggered starbursts in the outer regions, while the later coalescence triggers a stronger centralised starburst. 
We find non-axisymmetric features in the maps of burst mass fraction and dust attenuation in all galaxies, which could be caused by tidal effects during the recent merger. 
Comparisons with literature binary merger simulations suggests that the galaxies' rapid quenching was driven by gas consumption and the stabilisation against gas gravitational collapse by a growing spheroid, while AGN feedback was not necessarily a primary cause.
\end{abstract}




\section{Introduction} \label{sec:intro}
The shutting down of star formation in galaxies, known as “quenching”, is a crucial stage of galaxy evolution as galaxies transition from actively star-forming  into ``red and dead'' quiescent systems \citep[see][for a review]{delucia2025}. However, numerous physical mechanisms capable of driving quenching have been proposed in the literature, ranging from simple gas exhaustion, to energy injection from supernovae or central supermassive black holes, to influences from galaxy internal structure and the wider galaxy environment \citep[see][and references therein]{man2018}.  The relative contribution among these mechanisms remains unclear.

A common approach to constrain the relative importance of different quenching mechanisms is to study spatially resolved patterns in star formation. The simplest pattern identifies ``inside-out'' vs. ``outside-in'' quenching, where either the inner or outer region commences quenching first leading to radial gradients in current star formation rate (SFR), specific star formation rate  (sSFR = SFR/stellar mass), mean stellar ages or star formation histories \citep[e.g.][]{gonzalezdelgado2016, spindler2018, lin2019, bluck2020}.  Quenching mechanisms that are driven by centralised processes, such as active galactic nuclei (AGN) feedback and stellar/supernovae feedback from central bursts of star formation, are expected to result in inside-out quenching and positive SFR gradients  \citep{zubovas2016,ni2025,mcclymont2025}. On the other hand, environmental quenching mechanisms, such as starvation of gas, tidal effects and ram-pressure stripping, could result in gas being lost preferentially from the outer regions of galaxies leading to “outside-in” quenching \citep{schaefer2017,finn2018}. While direct observation of SFR gradients at different cosmic epochs or fossil record star formation history gradients in local galaxies can both help identify broad trends in quenching patterns, the long cosmic timescale over which galaxies grows means that multiple quenching mechanisms may be at play at different epochs of a galaxy’s life, complicating interpretation. Additionally, the signatures of inside-out quenching can easily be confused with those of inside-out growth \citep{lian2017,avilareese2023}.

An alternative approach is to study individual galaxies caught in the act of quenching. Post-starburst galaxies (PSBs) are of particular interest in this subject, as they have quenched their star formation $\lesssim1.5\;$Gyr before they were observed on timescales of a few 100\,Myr \citep[e.g.][]{du2010,melnick2013,french2018,wild2020,wu2023,paper1}. These galaxies' star formation histories (SFHs) lead their light to be dominated by A and F stars but lack contributions from shorter-lived O and B stars, resulting in weak or absent optical nebular emission \citep[see][for a recent review]{french2021_review}. Although they represent only $\sim1\,$ per cent of the galaxy population at $z<1$ \citep{goto2003,quintero2004,pawlik2016,pattarakijwanich2016,rowlands2018a}, their transient nature and increasing prevalence at higher redshift means they may contribute significantly to building up the total budget of quenched galaxies \citep{tran2004,wild2009,wild2016,whitaker2012,rowlands2018a,belli2019,forrest2020,taylor2023,park2023,skarbinski2026}, and may be a key stage within the cycles of stochastic SFH (``burstiness'') observed at very-high redshifts \citep{strait2023,looser2024,covelo-paz2025,hutchison2025}.

Recovering both the spatially and temporally resolved growth and quenching history of galaxies is crucial to fully understand the quenching processes. When compared to control galaxies matched in stellar mass and total SFR, most local PSBs generally have younger stellar populations in the centre than in the outer regions \citep{chen2019,deugenio2020,wu2021,cheng2024}. While the vast majority of PSBs have been identified in single fibre surveys, and are therefore by definition centrally located in the galaxy (central post-starbursts or CPSB), the advent of large integral field surveys (IFS) allowed post-starburst populations to be discovered for the first time in the outer regions of galaxies in which the inner regions are mostly still star-forming \citep[ring post-starbursts or RPSB,][]{chen2019}. Both of these results can be interpreted as evidence for outside-in quenching in PSBs, and a natural progression of RPSBs into CPSBs \citep{cheng2024,vasquezbustos2025}, in apparent contradiction to the expected inside-out quenching from a central starburst and/or growing supermassive black hole (SMBH). However, detailed star formation history analysis actually reveals true outside-in quenching in only a minority ($\sim10\,$ per cent) of RPSBs \citep{paper2}, with the gradient in stellar age driven instead by the greater strength of the preceding starburst in the central regions, or a slower rate of the decline in star formation in the central regions. This provides a concrete example of the complexities of disentangling the effects of radial gradients in both growth and quenching. 

Major gas-rich galaxy mergers are often suggested as a main driver for the characteristic recent SFH of PSBs \citep{bekki2005,wild2009,snyder2011,pawlik2019,davis2019,zheng2020,zanella2023}. Disruptions to the gravitational potential caused by mergers can lead to strong torques that funnel gas towards galaxy centres, which fuels a strong starburst \citep{barnes1991,barnes1996}. As the gas inflow halts while available gas is rapidly consumed by star formation or removed through feedback, star formation can rapidly quench. Turbulence in the ISM caused by AGN feedback \citep{piotrowska2022}, disturbances to the gravitational potential following the merger \citep{scoville2017} or a rotating bar \citep{salim2020b} could also hinder star formation even when gas is readily available, which may be a key local-scale process that regulates the prolonged quiescence following quenching in PSBs \citep{smercina2018,otter2022,paper2}. 

Galaxy mergers as triggers for PSBs is supported by various observational evidence, where local PSBs are more likely to show morphological signatures of recent mergers than controls, particularly at high stellar masses \citep[$>10^{10.3}$M$_\odot$, e.g.][]{pawlik2018,wilkinson2022, ellison2024}, and high fractions of post-mergers are found to host PSB populations  \citep{ellison2022,li2023,ellison2024}. Moreover, in \cite{paper1} and \cite{paper2}, we found the PSB regions in most local PSBs have formed significantly more metal-rich stars during and after the starburst than before, which could be the result of gas-compression-triggered starbursts following recent mergers as indicated by hydrodynamical simulations \citep{perez2011,torrey2012}. There remains some question about the origin of lower mass PSBs in the low-redshift Universe in particular: often no merger signatures are seen \citep{pawlik2018,ellison2024}, yet some process must have driven a short intense burst of star formation to form the spectral features. Minor and micro mergers are leading candidates \citep{pawlik2019}, alongside feedback from SNe, environment \citep{sandovalascencio2025} or AGN \citep{wong2012}.

The strong link between PSBs and galaxy mergers is pertinent to understanding galaxy quenching, because it is still widely debated to what extent galaxy mergers are responsible for quenching of star formation in the Universe. The physical mechanisms responsible for star formation quenching following galaxy mergers are also unclear, with hydrodynamic simulations implicating gas exhaustion \citep{barnes1992,wild2009}, feedback from the central SMBH \citep[e.g.][]{hopkins2013, zheng2020,sanchez2021, davies2022,quai2023}, and stabilisation of the gas to collapse \citep[morphological quenching or dynamical suppression, ][]{martig2009, gensior2020, gensior2021, petersson2023, porter2026} as the dominant causes. Even the physical origin of merger-induced star formation is not fully understood \citep{chienbarnes2010,renaud2019,moreno2021,petersson2023}, complicating our ability to understand subsequent quenching mechanisms. Finally, a lack of agreement between simulations and observations further restricts efforts to understand the true impact of galaxy mergers on the quenching of star formation in galaxies \citep{zheng2022,quai2023}.

In this paper we aim to answer two outstanding questions: is the spatially resolved star formation history of local post-starburst galaxies 
\begin{enumerate}
    \item consistent with them being caused by a gas-rich major merger?
    \item a strong constraint on the dominant physical processes responsible for star formation quenching?
\end{enumerate}

In \citet{paper2} we demonstrated that simple stellar age-sensitive spectral indices provide insufficient temporal resolution to properly map local quenching patterns, and full SFH modelling is necessary. In this paper we further expand on this step-change in methodology, by developing a robust hierarchical Bayesian fitting method that measures the full spatially resolved SFH of PSBs, while simultaneously modelling whole-galaxy characteristics such as radial gradients and stellar mass profiles. This approach enables the inclusion of a-priori knowledge about galaxy structures and correlated noise properties of the data, to improve constraints on the key physical properties of interest such as the starburst age, strength and rate of decline, particularly in lower surface-brightness outer regions. These regions are crucial for investigating the physical processes responsible for starbursts and quenching in different galactic regions. Hierarchical modelling of spatially resolved galaxy properties has been done before on multi-band photometry \citep{sanchez-gil2019}, and IFS data \citep{gonzalez-gaitan2019,ditrani2024}, showcasing the method's capability in revealing detailed patterns out to large radii in galaxies of interest.

The structure of this paper is as follows. 
In Section \ref{sec:data}, we describe the data and our choice of galaxies from the MaNGA survey. Section \ref{sec:theory} lays down the statistical theory of our hierarchical method and provides examples of its application in astronomy, while Section \ref{sec:Hbayes_methods} details our model assumptions, setup, priors and the sampling procedure. Section \ref{sec:theory} can be safely skipped for readers less interested in the detailed methodology. The measured radial gradients and property maps of three extended PSBs are presented in Section \ref{sec:results}, and their implications for the origin of PSBs are discussed in Section \ref{sec:discussion}. 
Where necessary, we assume a cosmology with $\Omega_M=0.3$, $\Omega_\Lambda=0.7$ and $h = 0.7$.
All magnitudes are in the AB system \citep{oke1983}.
We assume a \citet{kroupa2001} stellar initial mass function (IMF), and take solar metallicity as $Z_\odot=0.0142$ \citep{asplund2009}.

\section{Data} \label{sec:data}
\begin{table*}
    \centering
    \caption{The basic properties and binning configurations of the three post-starburst galaxies in this study:
    (1) MaNGA Plate-IFU identifier;
    (2) MaNGA identifier;
    (3) R.A. (J2000);
    (4) Declination (J2000);
    (5) Redshift;
    (6) $\log_{10}$ total stellar mass fitted from elliptical Petrosian photometric fluxes in GALEX/SDSS \textit{FNugriz} bands from the NSA catalogue, adjusted for $h=0.7$;
    (7) Half light radius (along major axis) from 2D Sersic $r$-band fits from the NSA catalogue, adjusted for $h=0.7$;
    (8) Threshold on $\mathrm{SNR}_{g}$ during Voronoi binning;
    (9) Number of Voronoi bins.}
    \begin{tabular}{p{1.5cm}p{1.3cm}p{1.5cm}p{1.7cm}p{1.3cm}p{1.6cm}p{1.3cm}p{1.5cm}p{1.1cm}}
      \hline
      Plate-IFU (1) & MANGA ID (2) & RA (degrees) (3) & Dec. (degrees) (4) & Redshift (5) & $\log_{10}(M_*/\mathrm{M_\odot})$ (6) & $R_e / \mathrm{kpc}$ (7) & $\mathrm{SNR}_{g}$ threshold (8) & $N_\mathrm{bins}$ (9) \\
      \hline
      7965-1902   & 1-635485  & 318.50227 & 0.53509   & 0.0269    & 10.41 & 2.00 & 10 & 391 \\
      12067-3701  & 1-26470   & 332.02545 & -0.90695  & 0.0379    & 10.4 & 1.59 & 10 & 222 \\
      12514-3702  & 1-66669   & 198.46832 & 2.13256   & 0.0302    & 10.57 & 4.65 & 25 & 166 \\
      \hline
    \end{tabular}
    \label{tab:data}
\end{table*}

For this initial study we select three galaxies with high quality data from the sample of 50 local galaxies with central PSB regions studied in \cite{paper1}. Their Plate-IFU identifiers are 7965-1902, 12067-3701 and 12514-3702 (selection explained below). The data is taken from the integral field spectrograph MaNGA survey \citep{MANGA}, which is a part of the fourth generation Sloan Digital Sky Survey \citep[SDSS-IV][]{SDSS_IV}. MaNGA surveyed $\sim10000$ galaxies with $M_* > 10^9 M_\odot$ (11273 datacubes) in the local $z<0.2$ neighbourhood from 2014 to 2020. The spatial resolution attained is $0.5$\arcsec, and the spectral resolution is $R\simeq2000$ across wavelength range $3600 < \lambda < 10000\,${\AA}.

In this work, we aim to map the local SFH, chemical and dust properties of PSBs at the highest physical resolution possible. Therefore, to limit the impact of the point spread function (PSF) on smoothing out local variations in the spectra, we selected face-on galaxies from the 50 PSBs with large angular sizes and extensive, continuous PSB spaxels over a large portion of the galaxy. To minimize the complexity of the SFH in each spaxel and to maximize the variations in stellar populations between spaxels at different radial distances, we examine the PSBs' SDSS 3-colour image to remove galaxies with clear asymmetries that can potentially complicate the analysis. Lastly, we select three PSBs with the highest mean SNR of the stacked PSB spectrum in Table 1 of \cite{paper1}. Details of these galaxies are provided in Table \ref{tab:data}. Galaxy total stellar masses are taken from the NASA-Sloan Atlas (NSA; \texttt{NSA\_ELPETRO\_MASS}, see \citealt{blanton2011}).

\begin{figure}
    \centering
    \includegraphics[width=\columnwidth]{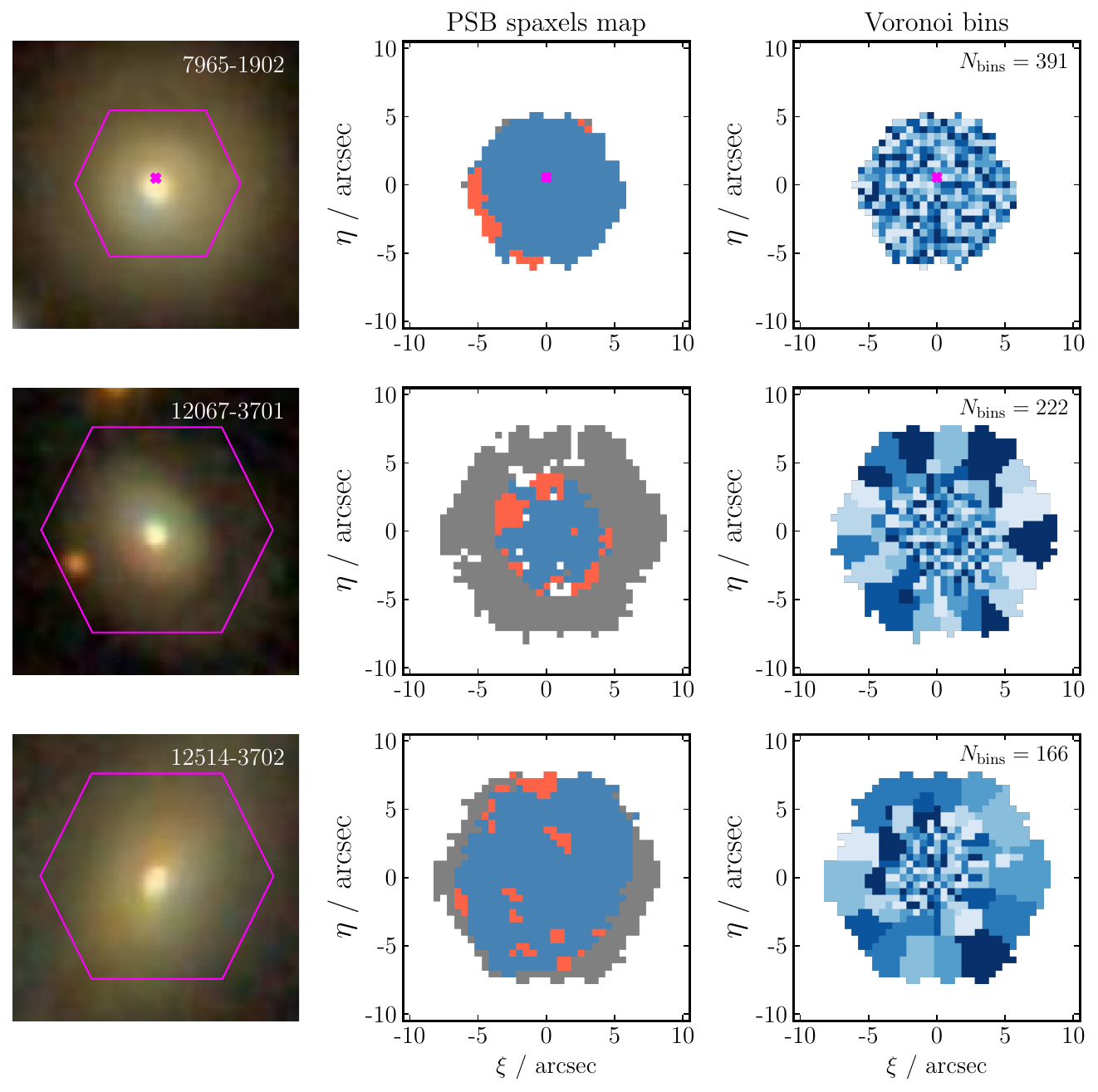}
    \caption{SDSS 3-colour images (\textbf{left}), the PSB spaxel selection (\textbf{middle}) and the Voronoi bin distribution (\textbf{right}) of our sample of three PSBs. Each galaxy's Plate-IFU is marked on the top right corner of the SDSS images. The MaNGA field of view is marked as the pink hexagon. In the middle column, we divide the spaxels into regions with no/faulty observations (not coloured), median spectral $\mathrm{SNR}<8$ too low to be classified (grey), classified as PSB (blue), and classified as non-PSB (red). Voronoi binning is performed with a threshold\footref{note1} $\mathrm{SNR}_{g}=10$ and accounting for spatial covariance. The colours in the right column are used to differentiate between spaxels in a given Voronoi bin. The magenta crosses in the top panels mark the location of the spaxel discussed in detail in Fig. \ref{fig:Bayes_vs_HBayes} and Section \ref{sec:discussion_hbayes}.}
    \label{fig:Hbayes_PSBmaps}
\end{figure}

We show the SDSS 3-colour images of the three chosen galaxies in the left column of Fig. \ref{fig:Hbayes_PSBmaps}. All three galaxies appear consistent with early-type morphologies and have prominent central bulges. 
Galaxy Zoo classified all three as ``smooth'' (no obvious spiral or disc features), and do not have additional irregular features \citep{willett2013}.
We include the morphological classifications of the three galaxies by \cite{vazquez-mata2025} based on visual inspection of images from SDSS and DESI Legacy Imaging Surveys in Table \ref{tab:morphology}. The authors classified 7965-1902 as S0 (lenticular), 12067-3701 as Sa and 12514-3702 as Sab. They also labelled 12514-3702 as having tidal features. Close inspection of the image of 12067-3701 reveals potentially asymmetric outer regions, however, these are faint and would need to be confirmed by e.g. a shape asymmetry measurement or machine learning classifier \citep{pawlik2016,wilkinson2022}. \cite{vazquez-mata2025} additionally measured the concentration, asymmetry and clumpiness of the galaxies (CAS, \citealt{conselice2003}). All three were measured to have low $\sim0.15$ asymmetry values, typical of early-type galaxies \citep{vazquez-mata2025}, indicating that they are not obviously post-merger. 12067-3701 alone has a high $S=0.47\pm0.17$ clumpiness value.

In the central column of Fig. \ref{fig:Hbayes_PSBmaps}, we also include maps of their spaxel classification (PSB, non-PSB and unclassified) according to the method described in \cite{paper1}\footnote{PSB spaxel selection: $\mathrm{SNR}>8$ per pixel, strength of the H$\delta$ Balmer absorption line after accounting for emission line infilling H$\delta_\mathrm{A}>3${\AA}, equivalent width of the H$\alpha$ nebular emission line after accounting for underlying absorption $\mathrm{W(H\alpha)}<10${\AA}, and $\log \mathrm{W(H\alpha)<0.23\times H\delta_A - 0.46}$.}. As per our selection, the galaxies have extensive, central and continuous PSB spaxels.

\begin{table}
    \centering
    \caption{The morphological properties of the three post-starburst galaxies in this study measured by \protect\cite{vazquez-mata2025}.}
    \begin{tabular}{lllll}
      \hline
      Plate-IFU & Type & C & A & S \\
      \hline
      7965-1902   & S0 & $3.6\pm0.4$ & $0.15\pm0.005$ & $0.10\pm0.12$ \\
      12067-3701  & Sa & $2.9\pm0.6$ & $0.13\pm0.00$ & $0.47\pm0.17$ \\
      12514-3702  & Sab & $4.3\pm0.3$ & $0.14\pm0.003$ & $0.08\pm0.12$ \\
      \hline
    \end{tabular}
    \label{tab:morphology}
\end{table}

To reduce estimation uncertainties and to limit the computational time spent on individual spaxels with low SNR (e.g. all spaxels at radial distance $>3.5''$ in 12067-3701 have median spectral $\mathrm{SNR}<8$ per pixel), we do not directly fit these spaxels with low SNR. Instead, we perform Voronoi binning of the MaNGA datacubes, with a threshold\footnote{\label{note1}$\mathrm{SNR}_g$ is the average SNR per wavelength channel weighted by the $g$-band response function \citep{manga_dap}.} $\mathrm{SNR}_{g}=10$ and accounting for spatial covariance following the procedure detailed in Section 6 of \cite{manga_dap}. For 12514-3702, due to the high number of $\mathrm{SNR}_{g}>10$ spaxels, excellent spatial resolution could already be achieved with a higher SNR threshold. Therefore, we instead opt for a threshold of $\mathrm{SNR}_{g}=25$ for this galaxy. To perform the binning, we used version 4.0.4 of the python package \texttt{mangadap} \citep{manga_dap}, whose binning procedure uses the Voronoi algorithm of \cite{cappellari2003}. The binning maps of the three galaxies are shown in the right column of Fig. \ref{fig:Hbayes_PSBmaps}, and summarized in Table \ref{tab:data}. Due to its higher SNR across the full datacube, the binning of 7965-1902 yielded a similar number of bins to the number of observed spaxels. For each Voronoi bin, we use the luminosity-weighted elliptical polar distance as the bin's radial distance from the galaxy centre. Note that the binned spectra are the simple average spectrum from the spaxels covered by the bin, not the summed spectrum.

\section{Modelling context and theory} \label{sec:theory}
Suppose that we wish to model the observations of $N_V$ experimental units, each with associated data vector $\vec{D_i}$, and parameters $\vec{\theta_i}$ where $i=1, \dots, N_V$. In this study, the experimental units are spatially distinct local regions (Voronoi bins) of a single galaxy.
The $N_V$ regions have local properties that likely follow radial trends, which we would like to model. Therefore, we propose a hierarchical model with hyper-parameters $\vec{\alpha}$ that describe the radial dependence. The joint posterior distribution of $\vec{\alpha}$ and $\mathbf{\Theta}$ (all $\vec{\theta_i}$) is given by
\begin{equation}
    P(\vec{\alpha},\mathbf{\Theta}|\mathbf{D}) = \frac{P(\mathbf{D}|\mathbf{\Theta},\vec{\alpha}) \; P(\mathbf{\Theta},\vec{\alpha})}{P(\mathbf{D})} \; ,
\end{equation}
where $P(\mathbf{D}|\mathbf{\Theta},\vec{\alpha})$ is the likelihood, $P(\mathbf{\Theta},\vec{\alpha})$ is the prior and $P(\mathbf{D})$ is the evidence. $\mathbf{D}$ denotes all $\vec{D_i}$ for all $N_V$ regions.

As we show in Appendix \ref{apx:derivations1}, the joint posterior distribution can be rewritten as a product over $i$:
\begin{equation} \label{eq:Hbayes_main}
    P(\vec{\alpha},\mathbf{\Theta}|\mathbf{D}) = \frac{P(\vec{\alpha}) \prod^{N_V}_{i=1}\Big[P(\vec{D_i}|\vec{\theta_i}) \; P(\vec{\theta_i}|\vec{\alpha})\Big]}{P(\mathbf{D})} \; ,
\end{equation}
where $P(\vec{\alpha})$ is the hyper-prior, $P(\vec{D_i}|\vec{\theta_i})$ and $P(\vec{\theta_i}|\vec{\alpha})$ are the likelihood and conditional prior for the $i$-th region, respectively.

\subsection{The practical case}
Our goal is to estimate the unknown $\mathbf{\Theta}$ and $\vec{\alpha}$ by sampling from the joint posterior distribution $P(\vec{\alpha},\mathbf{\Theta}|\mathbf{D})$ given by Equation \ref{eq:Hbayes_main}. However, direct sampling from this distribution is not straightforward. The number of unknown parameters to be constrained (dimensionality) in the hierarchical model is $N_V \times l_\theta + l_\alpha$ where $l_\theta$ is the number of parameters in $\vec{\theta_i}$ and $l_\alpha$ is the number of hyper-parameters. Depending on $N_V$, this can easily reach $>100$ or even $>1000$ dimensions, significantly higher than non-hierarchical Bayesian models. Direct sampling from this extremely high dimensional posterior volume can be challenging.

Therefore, we perform our fitting through a three-stage process that avoids direct sampling from the hierarchical model:
\begin{enumerate}
    \item Fit each Voronoi bin independently with a non-hierarchical model (Section \ref{sec:theory_stage1})
    \item Obtain posterior samples of hyper-parameters from the hierarchical model by sampling from the marginalised posterior $P(\vec{\alpha}|\mathbf{D})$, applying principles of importance sampling (Section \ref{sec:theory_stage2})
    \item Perform rejection sampling to obtain samples from $P(\vec{\theta_i}|\vec{\alpha})$. The combined samples for $\vec{\alpha}$ and $\vec{\theta_i}$ are the samples from the posterior in Equation \ref{eq:Hbayes_main}, i.e. the one implied by the assumed hierarchical model, as they were drawn from the collapsed portion of a converged partially collapsed Gibbs sampler \citep[see Sampler 3 in][]{van_dyk2008} (Section \ref{sec:theory_stage3})
\end{enumerate}
Sampling in all stages are performed with the neural-network-boosted nested sampling algorithm \textsc{nautilus} \citep{nautilus}.

\subsubsection{Stage 1} \label{sec:theory_stage1}
First, we perform sampling in all Voronoi bins from a non-hierarchical model. For the $i$-th Voronoi bin, the posterior samples obtained are drawn from the posterior distribution
\begin{equation} \label{eq:theory_non_hierarchical}
    P(\vec{\theta_i^\mathrm{NH}}|\vec{D_i}) = \frac{P(\vec{D_i}|\vec{\theta_i^\mathrm{NH}}) \; P(\vec{\theta_i^\mathrm{NH}})}{P(\vec{D_i})} \; ,
\end{equation}
where $\vec{\theta_i}$ is the vector of all model parameters that describes the model spectrum of the $i$-th Voronoi bin, and $\vec{D_i}$ is the observed binned spectrum of the same bin. Non-hierarchical parameters are denoted with the superscript ``NH''. In practice, sampling is performed on the unnormalised posterior distribution, where the $P(\vec{D_i})$ term in the denominator is ignored.

\subsubsection{Stage 2} \label{sec:theory_stage2}
Given the samples from the non-hierarchical models obtained in \textit{stage 1}, we can apply principles of importance sampling to obtain samples from the distribution $P(\vec{\alpha}|D)$ implied by the hierarchical model. It is similar to the sampling methods applied in \cite{hogg2010}, \cite{forman-mackey2014}, and \cite{thrane2019}. 

From the hierarchical model, the marginalised posterior $P(\vec{\alpha}|D)$ where all $\mathbf{\Theta}$ are integrated out is given by
\begin{align} \label{eq:integral}
    P(\vec{\alpha}|\mathbf{D}) &= \int P(\vec{\alpha},\mathbf{\Theta}|\mathbf{D}) \; d\mathbf{\Theta} \\
    &= \frac{P(\vec{\alpha})}{P(\mathbf{D})} \prod^{N_V}_{i=1} \int P(\vec{D_i}|\vec{\theta_i}) \; P(\vec{\theta_i}|\vec{\alpha}) \; d\vec{\theta_i} \; .
    \label{eq:theory_1}
\end{align}

Notice that the likelihood $P(\vec{D_i}|\vec{\theta_i},\vec{\alpha})$ is equal to $P(\vec{D_i}|\vec{\theta_i})$ in Equation \ref{eq:theory_1}, as $\vec{D_i}$ is independent of $\vec{\alpha}$ conditionally on $\vec{\theta_i}$. Applying the principles of importance sampling, we can obtain samples of $\vec{\theta_i}$ drawn from $P(\vec{\theta_i}|\vec{\alpha})$ through sampling from some other PDF of $\vec{\theta_i}$. In this case, the other PDF is $P(\vec{\theta_i^\mathrm{NH}}|\vec{D_i})$ in Equation \ref{eq:theory_non_hierarchical} obtained from the non-hierarchical model. 
We multiply top and bottom of the integrand in Equation \ref{eq:theory_1} by $P(\vec{\theta_i^\mathrm{NH}}|\vec{D_i})$ and substitute Equation \ref{eq:theory_non_hierarchical} to give
\begin{equation} \label{eq:theory_4}
    P(\vec{\alpha}|D) = \frac{P(\vec{\alpha})}{P(\mathbf{D})} \prod^{N_V}_{i=1} \int \frac{P(\vec{D_i}|\vec{\theta_i}) \; P(\vec{\theta_i}|\vec{\alpha}) \; P(\vec{D_i})}{P(\vec{D_i}|\vec{\theta_i^\mathrm{NH}}) \; P(\vec{\theta_i^\mathrm{NH}})} \; P(\vec{\theta_i^\mathrm{NH}}|\vec{D_i}) \; d\vec{\theta_i} \; .
\end{equation}

The likelihood function of the hierarchical model $P(\vec{D_i}|\vec{\theta_i})$ is defined to be equal to the likelihood function of the non-hierarchical model $P(\vec{D_i}|\vec{\theta_i^\mathrm{NH}})$, which cancels out in Equation \ref{eq:theory_4}. We can also absorb the normalisation scalars $P(\mathbf{D})$ and $P(\vec{D_i})$ into a proportionality sign. The expression reduces to
\begin{equation} \label{eq:theory_5}
    P(\vec{\alpha}|D) \propto P(\vec{\alpha}) \prod^{N_V}_{i=1} \int \frac{P(\vec{\theta_i}|\vec{\alpha})}{P(\vec{\theta_i^\mathrm{NH}})} \; P(\vec{\theta_i^\mathrm{NH}}|\vec{D_i}) \; d\vec{\theta_i} \; .
\end{equation}

Posterior samples from the non-hierarchical model can be used to compute expectation values of quantities of interest, where the expectation value for a function $f(x)$ given the PDF $P(x)$ can be approximated by the mean of the function calculated at a representative number of samples drawn directly from the PDF:
\begin{equation}\label{eq:exp_val}
    \langle f(x)\rangle_{P(x)} = \int dx P(x) f(x) \approx \frac{1}{n_s}\sum^{n_s}_{j=1}f(x_j) \; ,
\end{equation}
where the sum over $j$ runs over $n_s$ samples \citep[e.g.][]{hogg2018}. From Equation \ref{eq:theory_5}, substituting $P(\vec{\theta_i^\mathrm{NH}}|\vec{D_i})$ as $P(x)$, and $\frac{P(\vec{\theta_i}|\vec{\alpha})}{P(\vec{\theta_i^\mathrm{NH}})}$ as $f(x)$, using Equation \ref{eq:exp_val} we have
\begin{equation} \label{eq:theory_hierarchical}
    P(\vec{\alpha}|D) \propto P(\vec{\alpha}) \prod^{N_V}_{i=1} \frac{1}{n_{s,i}} \sum^{n_{s,i}}_{j=1} \frac{P(\vec{\theta_{i,j}}|\vec{\alpha})}{P(\vec{\theta_{i,j}^\mathrm{NH}})} \; ,
\end{equation}
which we can use to sample from the posterior distribution of $P(\vec{\alpha}|D)$. The conditional probability $P(\vec{\theta_{i,j}}|\vec{\alpha})$ is the probability of drawing the posterior sample values for $\vec{\theta_i}$ given by the $j$th posterior sample of the $i$th Voronoi bin, given the hyper-priors. 
The probability $P(\vec{\theta_{i,j}^\mathrm{NH}})$ is the probability of drawing $\vec{\theta_{i,j}}$ from the uninformative prior assumed during \textit{stage 1}. At the limit of an infinite number of samples $n_{s,i}$, Equation \ref{eq:theory_hierarchical} is identical to Equation \ref{eq:theory_4}, up to proportionality.

It is useful to make a further distinction in $\vec{\theta}$ between parameters that directly depend on the hyper-parameters ($\vec{\phi}=[M_*,t_\mathrm{burst},Z_\mathrm{old},A_V,\sigma_\mathrm{disp}]$, see Section \ref{sec:Hbayes_stage2}) and parameters that do not ($\vec{\nu}$). For $\vec{\nu}$, their conditional prior $P(\vec{\nu_{i,j}}|\vec{\alpha})$ reduces to a normal prior that we assign. If we assign the same prior for these parameters in the hierarchical model as the priors in the non-hierarchical model in \textit{stage 1}, the term within the sum in Equation \ref{eq:theory_hierarchical} for these parameters reduces to 1. Therefore, we can rewrite the equation as
\begin{equation}
    P(\vec{\alpha}|D) \propto P(\vec{\alpha}) \prod^{N_V}_{i=1} \frac{1}{n_{s,i}} \sum^{n_{s,i}}_{j=1} \frac{P(\vec{\phi_{i,j}}|\vec{\alpha})}{P(\vec{\phi_{i,j}^\mathrm{NP}})} \; .
\end{equation}

All caveats of importance sampling also apply here: for the estimated posterior to have proper support, the limits of the uninformative priors must be wide enough to cover the range in $\vec{\theta_i}$ where the conditional probability $P(\vec{\theta_i}|\vec{\alpha})$ is non-zero for all values of hyper-parameters $\vec{\alpha}$ within its hyper-priors. 
Either a large number of samples of $\vec{\theta_{i,j}}$ must be drawn from the \textit{stage 1} or the two PDFs in the importance weights in Equation \ref{eq:theory_hierarchical} must be sufficiently similar to limit the noise of the importance sampling approximation.

\subsubsection{Stage 3} \label{sec:theory_stage3}
Next, we obtain posterior samples for each region's parameters $\vec{\theta_i}$ in the hierarchical model from the conditional $P(\mathbf{\Theta}|\mathbf{D},\vec{\alpha})$, given values $\vec{\alpha}$ from \textit{stage 2}. 
Following \cite{van_dyk2008}, in particular Sampler 3, our posterior samples of $\vec{\alpha}$ drawn from $P(\vec{\alpha}|\mathbf{D})$ can be considered as the samples drawn from the collapsed portion of a converged partially collapsed Gibbs sampler, as long as the sampler in \textit{stage 2} has converged. Then, following \cite{van_dyk2008}, samples of $\mathbf{\Theta}$ drawn from $P(\mathbf{\Theta}|\mathbf{D},\vec{\alpha_k})$ where $\vec{\alpha_k}$ is the $k$-th posterior sample from \textit{stage 2}, will be the same as the posterior samples of $\mathbf{\Theta}$ drawn directly from the joint posterior $P(\vec{\alpha},\mathbf{\Theta}|\mathbf{D})$. Therefore, the samples drawn in \textit{stage 2} combined with the samples drawn in this stage are samples drawn from the posterior $P(\mathbf{\Theta},\vec{\alpha}|\mathbf{D})$.

We perform rejection sampling on the posterior samples drawn from the non-hierarchical model in \textit{stage 1}, for all Voronoi bins. We show in Appendix \ref{apx:derivations2} that for the $k$-th posterior sample of hyper-parameters from \textit{stage 2} ($\vec{\alpha_k}$) and the $i$-th Voronoi bin, the rejection sampling acceptance probability can be given by 
\begin{equation}
    P(\vec{\theta_i}\;\mathrm{is\;accepted}) = \frac{P(\vec{\phi_i}|\vec{\alpha_k})/P_n(\vec{\phi_i})}{\sup_{\vec{\phi_i}} \Big[ P(\vec{\phi_i}|\vec{\alpha_k}) / P_n(\vec{\phi_i}) \Big] } \; . \label{eq:acceptance_prob}
\end{equation}
The samples drawn from this rejection sample process are samples drawn from $P(\vec{\theta_i}|\vec{D_i},\vec{\alpha})$. By repeating this process for each hyper-parameter posterior sample $\vec{\alpha_k}$, and combining the samples of $\vec{\theta_i}$ obtained, we construct a posterior sample the same as one drawn directly from the distribution $P(\mathbf{\Theta}|\mathbf{D})$ of the hierarchical model, marginalised across all $\vec{\alpha}$. 

\section{Fitting spatially resolved spectra with a Bayesian hierarchical model} \label{sec:Hbayes_methods}
To measure the spatially resolved SFHs, metallicity and dust properties of our three PSBs, we perform Bayesian hierarchical full spectral fitting on the Voronoi binned spectra described in Section \ref{sec:data}. To fit individual Voronoi bins, we broadly follow the fitting procedure and priors in \cite{paper1} designed to fit stacked spectra of all PSB regions in PSBs. The model parameters and priors are listed in Table \ref{tab:priors}. Details of this base model is described in Section \ref{sec:Hbayes_stage0} below. 

\begin{figure*}
    \centering
    \includegraphics[width=0.9\textwidth]{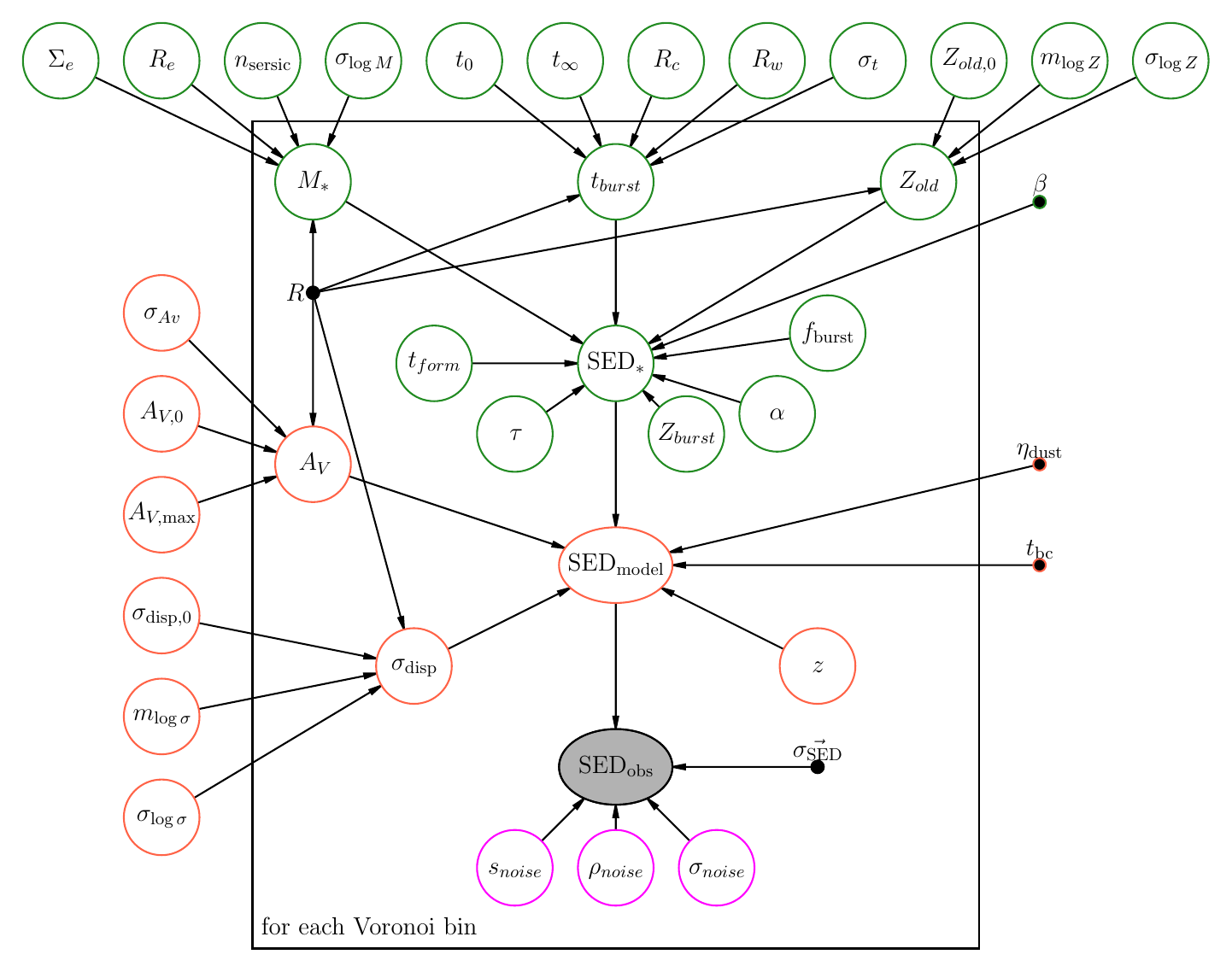}
    \caption{The probabilistic graphical model of the full hierarchical model. Open, shaded and closed nodes correspond to unknown (free), observed and fixed parameters, respectively. Nodes inside the rectangle are parameters unique to each individual Voronoi bins (Section \ref{sec:Hbayes_stage1}). Nodes outside the rectangle are hyper-parameters of the population model (Section \ref{sec:Hbayes_stage2}). The stellar portion of the model is marked with green rims, where the population model describes radial dependencies for the local stellar mass ($M_*$), burst age ($t_\mathrm{burst}$), and the pre-burst metallicity ($Z_\mathrm{old}$). Parameters related to dust attenuation, velocity dispersion and redshift are marked with orange rims, where the population model describes radial dependencies for the $V$-band dust attenuation strength ($A_V$) and velocity dispersion ($\sigma_\mathrm{disp}$). The additive Gaussian Process (GP) noise component is marked with magenta rims. Each Voronoi bin's $M_*$, $t_\mathrm{burst}$, $Z_\mathrm{old}$, $A_V$ and $\sigma_\mathrm{disp}$ have dependence on the bin's radial distance from the galaxy centre ($R$).}
    \label{fig:Hbayes_pgm}
\end{figure*}

For each galaxy, we introduce an additional layer of modelling that describes the Voronoi bin regions as a population, making our Bayesian model hierarchical. Our population model assumes that some of the properties (regional stellar mass, time since the recent starburst, pre-burst stellar metallicity, $A_V$ and $\sigma_\mathrm{disp}$) vary as a function of the region's radial distance to the galaxy's centre; see Section \ref{sec:Hbayes_stage1} for more details. A probabilistic graphical model visualization of our full hierarchical model is provided in Fig. \ref{fig:Hbayes_pgm}.

Additionally, we perform fitting on the globally stacked spectrum of each galaxy before \textit{stage 1} to construct more informative priors for later stages. This is labelled \textit{stage 0}. Sampling of the posterior surface in all stages is performed with the neural-network-boosted nested sampling algorithm \textsc{nautilus} \citep{nautilus}.

\subsection{Stage 0: Constructing more informative priors from fitting the globally stacked spectrum} \label{sec:Hbayes_stage0}
\begin{table*}
\centering
	\caption{Model priors used in \textit{stage 0} and \textit{stage 1} fitting globally stacked spectra and Voronoi bins individually. The parameter symbols are described in Section \ref{sec:Hbayes_stage0}, or otherwise have their usual meanings. \textbf{Global priors}: Prior shape $\log_{10}$ uniform indicates a flat prior in uniform space $\log(X) \sim U(\log(min), \log(max))$. The redshift prior is uniform ranging from 80 per cent to 120 per cent of the target's MaNGA redshift ($z$). \textbf{Individual priors}: Prior shape triple Gaussian is an approximation of the parameter's posterior distribution from the galaxy's global fit. $t_\mathrm{burst,84}$ is the 84-th percentile value of the global fit's $t_\mathrm{burst}$ posterior distribution. The upper limit of $t_\mathrm{burst}$ is 2 Gyr or $t_\mathrm{burst,84}+1.5\;$Gyr, whichever one is higher, with an upper prior limit of 4 Gyr. The dust birthcloud factor ($\eta$) is fixed at the maximum a posteriori (MAP) value of the global fit. Redshift is given a uniform prior centred at the MAP value of the global fit, with a range equivalent to $1000\;\mathrm{km/s}$ difference in line of sight velocity on both sides of the MAP value. Note that $\sigma_{\rm disp}$ is not the intrinsic velocity dispersion, as it does not account for the finite resolution of the spectral templates or instrument.}
	\label{tab:priors}
    \begin{tabular}{lllll|lll}
        \hline
        \multirow{2}{*}{Type} & \multirow{2}{*}{Parameter} & \multicolumn{3}{c|}{Global} & \multicolumn{3}{c}{Individual} \\  
         &  & Form & Min & Max & Form & Min & Max \\ \hline
        SFH & $\log_{10}(M_*/M_\odot)$ & Uniform & 6 & 13 & \multicolumn{3}{l}{Same as global} \\
         & $t_{\rm{form}}$ / Gyr & Uniform & 4 & 14 & Triple Gaussian & 4 & 14 \\
         & $\tau_e$ / Gyr & Uniform & 0.3 & 10 & Triple Gaussian & 0.3 & 10 \\
         & $t_{\rm{burst}}$ / Gyr & Uniform & 0 & 4 & Uniform & 0 & $(t_\mathrm{b,84}+1.5)<4$ or 2 \\
         & $\alpha$ & $\log_{10}$ Uniform & 0.01 & 1000 & \multicolumn{3}{l}{Same as global} \\
         & $\beta$ & Fixed = 250 & - & - & \multicolumn{3}{l}{Same as global} \\
         & $f_{\rm{burst}}$ & Uniform & 0 & 1 & \multicolumn{3}{l}{Same as global} \\
        Metallicity & $Z_{\rm{old}}/Z_\odot$ & $\log_{10}$ Uniform & 0.014 & 3.52 & \multicolumn{3}{l}{Same as global} \\
         & $Z_{\rm{burst}}/Z_\odot$ & $\log_{10}$ Uniform & 0.014 & 3.52 & \multicolumn{3}{l}{Same as global} \\
        Dust & $A_V$ / mag & Uniform & 0 & 2 & \multicolumn{3}{l}{Same as global} \\
         & birthcloud factor $\eta$ & Uniform & 1 & 5 & Fixed = MAP & - & - \\
         & $t_{\rm{birth cloud}}$ / Gyr & Fixed = 0.01 & - & - & \multicolumn{3}{l}{Same as global} \\
        GP noise & uncorrelated amplitude $s$ & $\log_{10}$ Uniform & 0.1 & 10 & \multicolumn{3}{l}{Same as global} \\
         & correlated amplitude $\sigma$ & $\log_{10}$ Uniform & $10^{-4}$ & 1 & \multicolumn{3}{l}{Same as global} \\
         & period/length scale $\rho$ & $\log_{10}$ Uniform & 0.04 & 1.0 & Triple Gaussian & 0.04 & 1.0 \\
         & dampening quality factor $Q$ & Fixed = 0.49 & - & - & \multicolumn{3}{l}{Same as global} \\
        Miscellaneous & redshift & Uniform & $0.8\;z$ & $1.2\;z$ & Uniform & MAP$-\sigma$ & MAP$+\sigma$ \\
         & $\sigma_{\rm{disp}}$ / km/s & $\log_{10}$ Uniform & 40 & 4000 & \multicolumn{3}{l}{Same as global} \\ \hline
    \end{tabular}
\end{table*}

Due to the lack of prior knowledge on the galaxies of interest, the priors used during Bayesian spectral fitting typically span multiple orders of magnitude \citep[e.g.][]{bagpipes2019,paper1}. This is designed to cover a parameter space that describes a broad range of SFH, chemical and dust properties. However, due to their uninformative nature, a large difference exists between posterior and prior volumes, thus reducing sampling efficiency.

For the purpose of fitting resolved spectra in a galaxy, we can safely assume that some parameters vary little spatially, thus can be given more informative priors. To do this, we first fit the globally stacked spectrum of the whole galaxy\footnote{For each galaxy, the globally stacked spectrum is constructed by an unweighted sum of all spaxels in the datacube that do not have quality flags \texttt{DEADFIBER} or \texttt{FORESTAR} in MaNGA's H$\alpha$ emission line maps. Uncertainties are summed in quadrature.}. From the global results, we then design more informative priors for fitting individual Voronoi bins in Section \ref{sec:informative_priors}.

\subsubsection{Fitting the global spectrum} \label{sec:global_fitting}
To perform full spectral fitting, we use the Bayesian fitting code \textsc{Bagpipes} \citep{bagpipes2018,bagpipes2019}, which is based on the \cite{bruzual2003} stellar population synthesis models (2016 version), and assumes the initial mass function from \cite{kroupa2001}. We broadly follow the methods in \cite{paper1}. Here, we provide a brief summary. The priors of the global fit are listed in the left of Table \ref{tab:priors}.

For the SFH model, we assume the two-component parametric SFH model for PSBs from \cite{wild2020}: 
\begin{equation}\label{eq:psb2}
    \mathrm{SFR}(t) \propto \frac{1-f_{\mathrm{burst}}}{\int \psi_e \mathrm{d}t} \times \psi_e(t)\Big|_{t_{\mathrm{form}}>t>t_{\mathrm{burst}}} 
    + \frac{f_{\mathrm{burst}}}{\int \psi_{\mathrm{burst}} \mathrm{d}t} \times \psi_{\mathrm{burst}}(t) \; .
\end{equation}
The SFH model includes an older, exponential decay component that models the existing stellar populations pre-starburst ($\psi_e$), and a younger, double power-law component that models the recent starburst and its quenching ($\psi_\mathrm{burst}$). Both components are functions of lookback time $t$. $t_{\mathrm{form}}$ is the lookback time when the older population began to form; $t_{\mathrm{burst}}$ is the time since the peak of the starburst; the fraction $f_{\mathrm{burst}}$ is the proportion of mass formed during the starburst. The two components are given by:
\begin{align}
\label{eq:exp}
    \psi_e(t') &= \exp^{\frac{-t'}{\tau_e}} \\
\label{eq:dpl}
    \psi_{\mathrm{burst}}(t') &= \Big[\big(\frac{t'}{t'_{\mathrm{burst}}}\big)^{\alpha_\mathrm{PSB}} 
    + \big(\frac{t'}{t'_{\mathrm{burst}}}\big)^{-\beta_\mathrm{PSB}}\Big]^{-1}\;,
\end{align}
where $t'$ is the age of the Universe, $t'_{\mathrm{burst}}$ is the age of the Universe at the peak of the starburst, $\tau_e$ is the e-folding timescale of the exponential component, while $\alpha_\mathrm{PSB}$ and $\beta_\mathrm{PSB}$ control the declining and increasing timescales of the burst respectively, with larger values corresponding to steeper slopes. Following \cite{wild2020}, \cite{paper1} and \cite{paper2}, we fix the rising slope of the starburst ($\beta=250$) and fit for all other parameters. 

To model any change in the metallicity of newly formed stars during the starburst, we assume the two-step evolutionary model from \cite{paper1}, which transitions near the peak of the starburst ($t=t_\mathrm{burst}$):
\begin{equation}
    Z(t) = \begin{cases}
    Z_{\mathrm{old}} & t>t_{\mathrm{burst}} \\
     Z_{\mathrm{burst}} & t\leq t_{\mathrm{burst}} \; .
    \end{cases}
\end{equation}
The pre-burst ($Z_{\mathrm{old}}$) and post-burst ($Z_{\mathrm{burst}}$) metallicity levels are mutually independent and have identical priors to not bias towards either rising or declining metallicity.

We apply the two-component dust attenuation law from \cite{wild2007} and \cite{dacunha2008}, with a fixed power-law exponent $n=0.7$ for the interstellar medium (ISM). The dust law asserts that stars younger than $10\,$Myr have a steeper power-law exponent $n=1.3$ and are more attenuated than older stars by a factor $\eta$ \citep[$=1/\mu$ in][]{wild2007,dacunha2008}, as they are assumed to be surrounded by their birth clouds.

To account for correlated observational uncertainties and model-data mismatch stemming from observational and calibrational issues and imperfect stellar templates, we employ a Gaussian Process (GP) noise model implemented through the \texttt{celerite2} python package \citep{celerite,celerite2} as an additive correctional term across the observed wavelength range. 

To ensure the fitted spectral wavelength range is fully within the MILES library's range used in the \cite{bruzual2003} models (2016 version) (\citealt{MILES}, see also \citealt{pawlik2018}), we limit all spectra to rest frame $\lambda < 7500\,${\AA}. We perform nested sampling with 2000 live points. To ensure our results are not biased by samples drawn during \textsc{Nautilus}' exploration phase due to the use of a pseudo-importance function, we follow \cite{nautilus} and discard all points obtained during the exploration phase.

\subsubsection{Designing more informative priors} \label{sec:informative_priors}
For later stages of fitting, we shrink or change the shape of 6 priors in Table \ref{tab:priors} given the results from the global fit. This change has the additional advantage of lowering computation time. We acknowledge that the data were used to a small extent to define the more informative  priors, however this is not using the data twice because we only use the overall structure and summary of the data to do this, given the stacking over all spatial dimensions to obtain the global spectrum. This practice is admissible within the Bayesian framework. One example is using the average of the observations to specify the prior mean of a log-linear model intercept, when adopting the $g$-prior for the model's parameters \citep[see][]{papathomas2018}. Our results below are not directly based on any of the parameters with updated priors. Additionally, since we only update the prior of generally poorly-constrained parameters, the new priors are not overly narrow. Therefore, we consider the risk of biasing our conclusions from this step to be small. The updated model priors used to fit the spectra from individual Voronoi bins are listed in the right of Table \ref{tab:priors}.

We assume that the SFH shape (not normalisation) of the stellar population formed before the starburst and the GP noise lengthscale ($\rho$) is similar across Voronoi bins. We perform least squares fits of a mixture of three normal distributions (the sum of three Gaussians each with varying amplitudes) to the 1D posterior distributions of three parameters: the time of beginning of star formation ($t_\mathrm{form}$), its exponential decay timescale ($\tau_e$) and $\rho$. The fitted mixture distributions are used as the priors for the spectral fitting of individual Voronoi bins. In cases where a single normal distribution alone can provide a good fit (if any of the three best fit Gaussian functions have an integrated area $>80\,$ per cent of the posterior distribution's area), we instead assume the simpler single normal distribution as the prior.

The birthcloud dust factor ($\eta$) is typically poorly constrained in our previous works with similar fitting models \citep{paper1,paper2}. Therefore, we fix $\eta$ in the fits of individual Voronoi bins as the maximum a posteriori (MAP) value from the galaxy's global fit.

Local stellar velocity along the line of sight can introduce spectral shifts in individual Voronoi bins. To model this, we allow the spectral shift of each bin to vary by $1000\;\mathrm{km/s}$ on both sides of the redshift MAP value from the galaxy's global fit, under a uniform prior.

Although the lookback time to the peak of the starburst ($t_\mathrm{burst}$) can vary between Voronoi bins at different radial distances, we do not expect large deviations from the global fit's posterior $t_\mathrm{burst}$. Therefore, we set the upper prior limit of $t_\mathrm{burst}$ for individual Voronoi bins in \textit{stage 1} as the higher value of 2 Gyr or $t_\mathrm{b,84}+1.5\;$Gyr, where $t_\mathrm{b,84}$ is the 84-th percentile value of the global fit's $t_\mathrm{burst}$ posterior distribution. We maintain the upper prior limit as 4 Gyr if $t_\mathrm{b,84}+1.5>4$.

\subsection{Stage 1: Fitting each Voronoi bin independently} \label{sec:Hbayes_stage1}
Using the same \textsc{Bagpipes} model as in Section \ref{sec:global_fitting}, we independently fit the binned spectrum from individual Voronoi bins of each galaxy using the updated priors listed in the right of Table \ref{tab:priors}. 

To decrease the sampling error in \textit{stage 2}, we must obtain a high number of posterior samples for all Voronoi bins at this stage. Therefore, we perform nested sampling for individual Voronoi bins with a minimum of 600 live points. If sampling efficiency is slow, we halt and restart fits with increased number of live points. As done in \textit{stage 0}, we discard all points obtained during the exploration phase. In the sampling phase that follows, we draw a minimum of 15000 weighted posterior samples for each Voronoi bin. This amounts to $\sim2000$ effective samples for the typical Voronoi bin.

\begin{figure*}
    \centering
    \includegraphics[width=\textwidth]{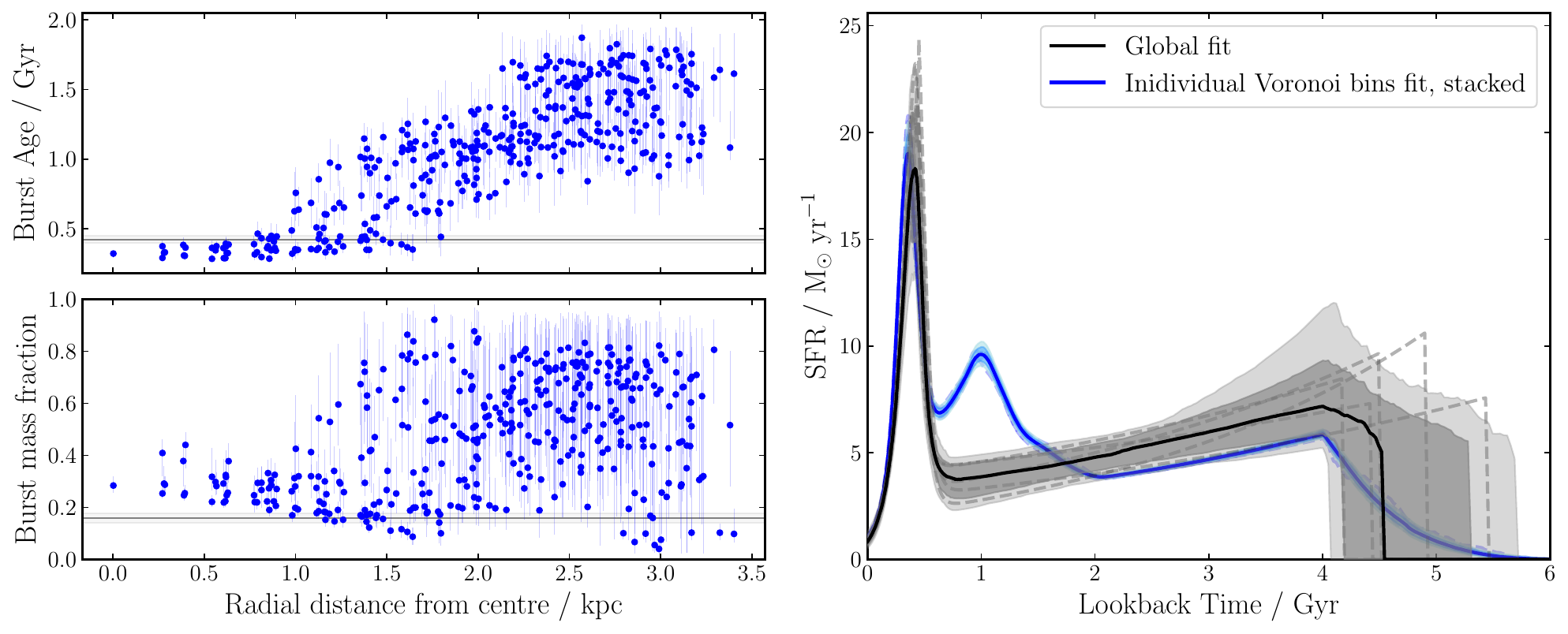}
    \caption{Fitting results of 7965-1902 from the non-hierarchical \textit{stage 0} and \textit{stage 1} models. The \textbf{left} panels plot the posterior burst age (top left) and burst mass fraction (bottom left) against radial distance from the galaxy centre for individual Voronoi bins in blue and the global fit as a gray horizontal line and shaded band. The \textbf{right} panel compares the SFH obtained from stacking the fitted SFH of all Voronoi bins (blue) to the fitted SFH from the global fit (black). The shaded regions denote 1 and $2\sigma$ uncertainty regions. The SFHs of 10 randomly drawn posterior samples are also shown as dashed curves. We observed the galaxy underwent two distinct starbursts, with the earlier one occurred predominantly in the outer regions, while the later, stronger one occurred predominantly in the centre. This is not captured by the global fit.}
    \label{fig:justify_tburst_gradient}
\end{figure*}

Here, it is of interest to compare results from the global fit from \textit{stage 0} and individual Voronoi bins, which can showcase the potential issues that arise when fitting spatially-integrated galaxy spectral energy distributions (SEDs).
The left panels in Fig. \ref{fig:justify_tburst_gradient} display the radial gradients in $t_\mathrm{burst}$ and $f_\mathrm{burst}$ of 7965-1902 measured from the non-hierarchical fits to individual Voronoi bins (this stage). The posterior median and $1\sigma$ uncertainty of the global fit is marked in the left panels as grey bands. At $R<1.0\;$kpc, the measured $t_\mathrm{burst}$ concentrates at $\sim0.4\;$Gyr, showing no obvious radial gradient. In the same radius range, a slight negative gradient in $f_\mathrm{burst}$ is visible. Between $R\simeq1.0\;$kpc and $R\simeq1.8\;$kpc, two distinct populations of Voronoi bins can be observed, where the lower $t_\mathrm{burst}$ population remains at $\sim0.4\;$Gyr and continues the negative $f_\mathrm{burst}$ gradient to lower values, while the higher $t_\mathrm{burst}$ population splits off to increase in both $t_\mathrm{burst}$ and $f_\mathrm{burst}$. At $R>1.8\;$kpc, the Voronoi bins are dominated by the high $t_\mathrm{burst}$ population, with a high $t_\mathrm{burst}\simeq1.4\;$Gyr and $f_\mathrm{burst}\simeq0.6$. 

In the right panel of Fig. \ref{fig:justify_tburst_gradient}, we stack the fitted SFHs from all Voronoi bins from the non-hierarchical fit (blue), and compare it to the global fitted SFH (grey). The stacked SFH from all Voronoi bins shows two distinct episodes of increased star formation within the last $\sim2\;$Gyr. The earlier, weaker episode at $\mathrm{lookback\ time}\sim1\;$Gyr is mainly contributed by the galaxy's outer regions, while the later, stronger episode at $\mathrm{lookback\ time}\sim0.4\;$Gyr that matches well with the SFH of the global fit's recent starburst is mainly contributed by the central region. This two-episode nature is used to inform the population model in \textit{stage 2}.

Comparing the two SFHs in Fig. \ref{fig:justify_tburst_gradient}, the global fit completely missed the earlier, weaker starburst at $\mathrm{lookback\ time}\sim1\;$Gyr that occurred in the galaxy's outskirts due to the outer region's lower luminosity than the centre. Therefore, one should be careful that when fitted spatially-integrated galaxy SEDs, as entire episodes of star formation can be missed.

\subsection{Stage 2: Modelling spatial patterns using radial dependence} \label{sec:Hbayes_stage2}

Based on known radial gradients in local galaxies in the literature and patterns we observed from the fitted parameters when performing the initial round of non-hierarchical spectral fitting in Section \ref{sec:Hbayes_stage1}, we design a population model that describes the radial trends of five resolved parameters in Table \ref{tab:priors}. The five parameters are the local stellar mass ($\log_{10}(M_*/M_\odot)$), burst age ($t_\mathrm{burst}$), pre-burst stellar metallicity ($Z_\mathrm{old}/Z_\odot$), dust attenuation in the ISM at the $V$-band ($A_V$) and stellar velocity dispersion ($\sigma_\mathrm{disp}$). Each radial model and their rationale are described in detail in the following sub-sections. To allow for azimuthal variations, we assume a Gaussian distribution centred at the population model value at the bin's radius for all five parameters. The priors of all hyper-parameters are given in Table \ref{tab:hyper_priors}. In total, the population model has 18 unknown (free) hyper-parameters.

\begin{table*}
    \centering
    \caption{Model hyper-priors of the population model used in \textit{stage 2} described in Section \ref{sec:Hbayes_stage2}.}
    \begin{tabular}{lllll}
        \hline
        Parameter & Hyper-parameter & Form & Min & Max \\ \hline
        Stellar mass formed & Effective radius $R_e$ / kpc & Uniform & 0.1 & 10 \\
         & $\log_{10}(\Sigma_e$ / $\mathrm{M_\odot\;kpc^{-2}})$ & Uniform & 6 & 13 \\
         & S\'ersic index $n$ & Uniform & 0.5 & 10 \\
         & Standard deviation $\sigma_{\log M}$ & Uniform & 0.0 & 0.1 \\
        Burst age & Burst age at $R=0$, $t_0$ / Gyr & Uniform & 0 & $(t_\mathrm{b,84}+1.5)<4$ or 2 \\
         & Burst age at $R=\infty$, $t_\infty$ / Gyr & Uniform & 0 & $(t_\mathrm{b,84}+1.5)<4$ or 2 \\
         & Inflection point radius $R_c$ / kpc & Uniform & 0 & $R_\mathrm{max}$ \\
         & Scale width $R_w$ / kpc & Uniform & 0 & $R_\mathrm{max}$ \\
         & Standard deviation $\sigma_t$ / Gyr & Uniform & 0 & 0.2 \\
        Pre-burst metallicity & Slope $m_{\log Z}$ / dex/kpc & Uniform & -0.1 & 0.1 \\
         & $Z_\mathrm{old}$ at $R=0$, $Z_{\mathrm{old},0}$ / $Z_\odot$ & $\log_{10}$ Uniform & 0.014 & 3.52 \\
         & Standard deviation $\sigma_{\log Z}$ / dex & Uniform & 0 & 0.05 \\
        Extinction in $V$-band & Extinction at $R=0$, $A_{V,0}$ / mag & Uniform & 0 & 2 \\
         & \begin{tabular}[c]{@{}l@{}}Extinction at $R=R_\mathrm{max}$, \\ $A_{V,\mathrm{max}}$ / mag\end{tabular} & Uniform & 0 & 2 \\
         & Standard deviation $\sigma_{A_V}$ / mag & Uniform & 0 & 0.4 \\
        Velocity dispersion & Slope $m_{\log \sigma}$ / dex/kpc & Uniform & -0.1 & 0.1 \\
         & $\sigma_\mathrm{disp}$ at $R=0$, $\sigma_{\mathrm{disp},0}$ / km/s & $\log_{10}$ Uniform & 40 & 4000 \\
         & Standard deviation $\sigma_{\log \sigma}$ / dex & Uniform & 0 & 0.1 \\ \hline
    \end{tabular}
    \label{tab:hyper_priors}
\end{table*}

\subsubsection{Local stellar mass} \label{sec:model_Mstar}
We adopt a S\'ersic profile \citep{sersic1963,sersic1968} to describe the intrinsic stellar mass surface density as a function of radius $R$. Note that instead of the more commonly measured surface density of existing stellar mass, here we model the surface density of the total stellar mass formed, which we refer to as the ``formed mass surface density'' for the rest of the paper. It can be written as
\begin{equation}
    \Sigma_*(R) = \Sigma_e \exp\bigg\{ -b_n \bigg[ \bigg(\frac{R}{R_e}\bigg)^{1/n}-1 \bigg] \bigg\}
\end{equation}
where $R_e$ is the half-mass radius, $n$ is the S\'ersic index, $\Sigma_e$ is the formed mass surface density at $R=R_e$, and $b_n$ is approximated as $b_n\simeq1.9992n-0.3271$ \citep{capaccioli1989}. We obtain the model formed mass surface density at each spaxel's radial distance, and multiply it with the datacube's spaxel physical size to yield the formed mass for each spaxel. 

MaNGA observations have a non-negligible point spread function (PSF) with a FWHM of $\sim\ang{;;2.54}$, which equates to a width of $\sim5$ spaxels \citep{manga_drp,yan2016b}. The PSF is well described by a 2D Gaussian profile and varies minimally with wavelength \citep{manga_drp}. Therefore, we convolve the formed mass grid with a 2D Gaussian kernel with a FWHM that matches the datacube's $r$-band PSF provided by the MaNGA data reduction pipeline. Finally, the model formed mass in each Voronoi bin is given by the mean\footnote{The mean instead of the sum is used here because the Voronoi binned spectra are the weighted average of all spaxels in the binned area, instead of the sum.} of all contributing spaxels. The three S\'ersic profile parameters $R_e$, $n$ and $\Sigma_e$ are set as unknown (free) hyper-parameters.

\subsubsection{Burst age} \label{sec:model_tburst}
\begin{figure}
    \centering
    \includegraphics[width=\columnwidth]{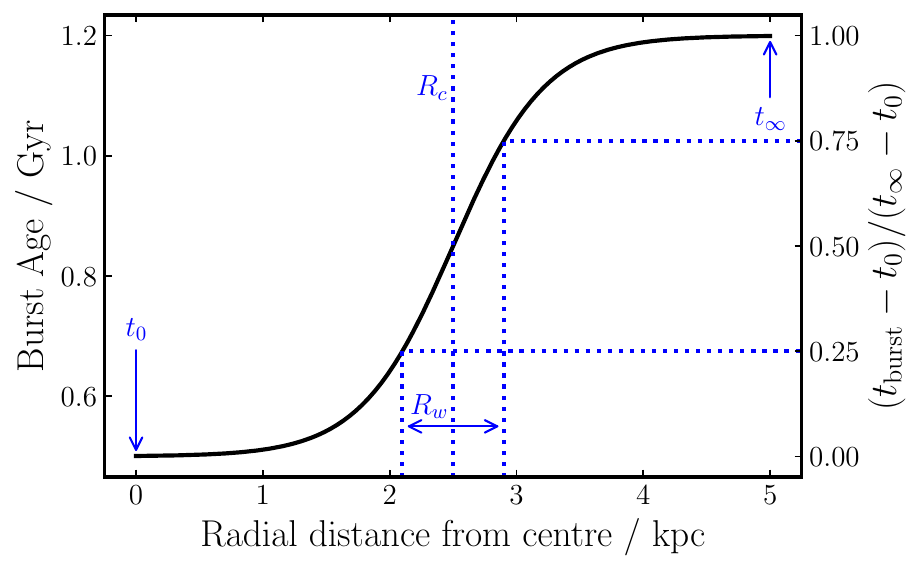}
    \caption{A visualization of the logistic function used to model the radial dependence of the galaxies' burst age in our population model. The right hand side y-axis shows a normalized view of the logistic function, which better determines the scale width $R_w$. For this example, the burst age at the centre is $t_0=0.5\;$Gyr, the burst age at the outer edge is $t_\infty=1.2\;$Gyr, the radial distance of the curve's inflection point is $R_c=2.5\;$kpc, and the scale width of the transition region is $R_w=0.8\;$kpc.}
    \label{fig:logistic_example}
\end{figure}

Initial non-hierarchical spectral fitting in Section \ref{sec:Hbayes_stage1} shows that the three PSB galaxies likely experienced two distinct starbursts at different radial distances. Therefore, we assume a modified logistic function to model the lookback time to the peak of the recent starburst ($t_\mathrm{burst}$) as a function of radius. A logistic function is well suited for the $t_\mathrm{burst}$ radial profile, since it asymptotes towards finite values at $x=-\infty$ and $x=\infty$. We use the functional form
\begin{equation}
    t_\mathrm{burst}(R) = \frac{t_\infty - t_0}{1+\exp[-k(R-R_c)]} + t_0 \; ,
\end{equation}
where $t_0$ is $t_\mathrm{burst}$ at $R=0$, $t_\infty$ is $t_\mathrm{burst}$ at $R\rightarrow\infty$, $R_c$ is $R$ at the central, inflection point of the logistic function, and $k=2\ln 3 / R_w$. $R_w$ is a scale width of the logistic function, defined as the radial distance between 1/4 and 3/4 of the difference between $t_0$ and $t_\infty$. Fig. \ref{fig:logistic_example} provides a visualization of the model. We set $t_0$, $t_\infty$, $R_c$ and $R_w$ as unknown (free) hyper-parameters. We assume identical priors for $t_0$ and $t_\infty$ such that both older or younger starbursts in the centre are equally likely given our priors.

\subsubsection{Pre-burst stellar metallicity} \label{sec:model_Zold}
Local massive PSBs are thought to arise from the mergers of gas rich disk galaxies \citep{pawlik2018,ellison2022,ellison2024}. \cite{paper1} showed that prior to the starburst, they follow the well known mass-metallicity relation. Local disk galaxies are known to have radial metallicity gradients \citep{roig2015,parikh2021,taibi2022}, which we model as a linear radial gradient in log metallicity:
\begin{equation}\label{eq:zmet_old_gradient}
    \log_{10}(Z_*/Z_\odot) (R) = m_{\log Z} R + \log_{10} Z_{\mathrm{old},0} \; ,
\end{equation}
where $m_{\log Z}$ is the slope of the radial gradient in log-space, and $Z_{\mathrm{old},0}$ is the pre-burst metallicity at $R=0$. $m_{\log Z}$ and $Z_{\mathrm{old},0}$ are set as unknown (free) hyper-parameters.

\subsubsection{Dust attenuation strength} \label{sec:model_Av}
Galaxies generally have a negative gradient in dust attenuation \citep{nelson2016,liu2017,greener2020}. We model the attenuation in the ISM with a linear radial gradient, given by
\begin{equation}
    A_V(R) = \frac{A_{V,\mathrm{max}} - A_{V,0}}{R_\mathrm{max}}R + A_{V,0} \; ,
\end{equation}
where $R_\mathrm{max}$ is the radial distance of the furthest Voronoi bin, $A_{V,\mathrm{max}}$ is the model $A_V$ at $R=R_\mathrm{max}$, and $A_{V,0}$ is the model $A_V$ at $R=0$. $A_{V,\mathrm{max}}$ and $A_{V,0}$ are set as unknown (free) hyper-parameters, while $R_\mathrm{max}$ is measured from the Voronoi bins.

\subsubsection{Stellar velocity dispersion} \label{sec:model_veldisp}
Most galaxies exhibit a negative or flat gradient in velocity dispersion \citep[e.g.][]{parikh2021}, which we model with a linear model in log velocity:
\begin{equation}
    \log_{10}(\sigma_\mathrm{disp}) (R) = m_{\log \sigma} R + \log_{10} \sigma_{\mathrm{disp},0} \; ,
\end{equation}
where $m_{\log \sigma}$ is the slope of the radial gradient in log-space, and $\sigma_{\mathrm{disp},0}$ is the velocity dispersion at $R=0$. Both are set as unknown (free) hyper-parameters.

\subsubsection{Modelling azimuthal variations}
To allow for local variations between Voronoi bins at the same radius, all 5 parameters with assumed radial dependence ($M_*$, $t_\mathrm{burst}$, $Z_*$, $A_V$ and $\sigma_\mathrm{disp}$) are drawn from a Gaussian distribution centred at the population model value at that bin's radial distance $R_i$, given by
\begin{align}
    \log_{10}M_{*,i} &\sim \mathcal{N}[\log_{10}(M_*(R_i)), \sigma_{\log M}] \\
    t_{\mathrm{burst},i} &\sim \mathcal{N}[t_\mathrm{burst}(R_i), \sigma_{t}] \\
    \log_{10}(Z_{*,i}/Z_\odot) &\sim \mathcal{N}[\log_{10}(Z_*/Z_\odot) (R_i), \sigma_{\log Z}] \\
    A_{V,i} &\sim \mathcal{N}[A_V(R_i), \sigma_{A_V}] \\
    \log_{10}(\sigma_{\mathrm{disp},i}) &\sim \mathcal{N}[\log_{10}(\sigma_\mathrm{disp}) (R_i), \sigma_{\log \sigma}] \; ,
\end{align}
where $\sigma_{\log M}$, $\sigma_{t}$, $\sigma_{\log Z}$, $\sigma_{A_V}$ and $\sigma_{\log \sigma}$ are the radius-independent standard deviations of their respective Gaussian distributions. All five standard deviations are set as unknown (free) hyper-parameters.

As detailed in Section \ref{sec:theory_stage2}, to obtain posterior samples of the hyper-parameters of the population model, we apply the principles of importance sampling given the posterior samples from the non-hierarchical model obtained in \textit{stage 1}. For all sampling of posteriors in \textit{stage 2}, we require 1000 live points. As done in previous stages, we discard all points obtained during the exploration phase.

\subsection{Stage 3: Rejection sampling to obtain the true posteriors for individual Voronoi bins} \label{sec:Hbayes_stage3}
As detailed in Section \ref{sec:theory_stage3}, we perform rejection sampling on the posterior samples drawn from the non-hierarchical model in \textit{stage 1} for each Voronoi bin given the posterior samples of hyper-parameters from \textit{stage 2}. This process yields posterior samples for all parameters (both the ones that we assume population models for in Section \ref{sec:Hbayes_stage2} and the ones that we do not) in all Voronoi bins in the hierarchical model.

\subsection{Assuming conditional independence between Voronoi bins}
One key assumption has been made in our hierarchical model's sampling framework. In order to split the sampling of the joint posterior distribution $P(\mathbf{\Theta},\vec{\alpha}|\mathbf{D})$ into the three separate stages, we have assumed that the Voronoi bins have no mutual dependence beyond what is described by our population model. This assumption allows for the conditional prior $P(\mathbf{\Theta}|\vec{\alpha})$ in Equation \ref{eq:Hbayes_theory4} to be separable into the product $\prod^{N_V}_{i=1}P(\vec{\theta_i}|\vec{\alpha})$, and the integral across all $\mathbf{\Theta}$ in Equation \ref{eq:integral} to be split. In reality this assumption is likely not valid due to the effects of the PSF. The PSF scatters light emitted from the galaxy region in one Voronoi bin into neighbouring Voronoi bins, thus introducing mutual dependence between Voronoi bins. We modelled the PSF effect in our population model's stellar mass component, but not for all other parameters. Hence, our observed radial gradients for properties other than stellar mass better reflect a PSF-convolved gradient than a true gradient.

In addition, due to the construction of the MaNGA datacubes from individual dithered observations, spatial covariance exists between neighbouring spaxels, and by extension, Voronoi bins \citep[see Section 9 in][]{manga_drp}. Therefore, our assumption of mutual independence between Voronoi bins is imperfect. This spatial covariance can be modelled by replacing the current 1D GP noise component with a 3D GP noise, including two spatial dimensions and one wavelength dimension. However, a 3D implementation would make it impossible to split the joint integral with respect to $d\mathbf{\Theta}$ in Equation \ref{eq:integral} into the product of many integrals with respect to $d\vec{\theta_i}$. 
This is a limitation of our method, as the splitting of the integral is necessary due to computational constraints. In the future, it may be possible to design machine learning methods to approximate the covariance of the dataset in a computationally reasonable time. Although we expect our conditional independence assumption to only have small effects on our results (it cannot change the sign of an observed gradient), we caution the reader to consider our results as estimations of the PSF-convolved gradients.

\renewcommand{\arraystretch}{1.4}
\begin{table}
    \centering
    \caption{Posterior estimated population properties (hyper-parameters) of the three CPSBs from spectral fitting of Voronoi binned MaNGA spaxels using hierarchical Bayesian model. The parameters have meanings and units as detailed in Table \ref{tab:hyper_priors}.}
    \begin{tabular}{lllll}
        \hline
        Plate-IFU & 7965-1902 & 12067-3701 & 12514-3702  \\
        \hline
        $R_e$                        & $3.65^{+0.81}_{-0.55}$            & $8.92^{+0.79}_{-1.25}$            & $3.15^{+0.76}_{-0.50}$             \\ 
        $\log_{10}\Sigma_e$          & $8.23^{+0.12}_{-0.15}$            & $7.40^{+0.11}_{-0.06}$            & $8.58^{+0.14}_{-0.17}$             \\ 
        $n_\mathrm{sersic}$          & $3.91^{+0.31}_{-0.27}$            & $5.15^{+0.16}_{-0.24}$            & $3.63^{+0.43}_{-0.35}$             \\ 
        $\sigma_{\log M}$            & $0.0840\pm0.0034$                 & $0.0996^{+0.0003}_{-0.0006}$      & $0.0952^{+0.0032}_{-0.0044}$       \\ 
        \hline
        $t_0$                        & $0.29^{+0.03}_{-0.04}$            & $0.98^{+0.03}_{-0.04}$            & $0.71^{+0.04}_{-0.06}$             \\ 
        $t_\infty$                   & $1.25^{+0.04}_{-0.03}$            & $1.89^{+0.07}_{-0.08}$            & $1.58^{+0.10}_{-0.09}$             \\ 
        $R_c$                        & $1.62^{+0.05}_{-0.04}$            & $2.37\pm0.09$                     & $1.98\pm0.04$                      \\ 
        $R_w$                        & $0.7\pm0.1$                       & $0.6\pm0.2$                       & $0.2^{+0.3}_{-0.1}$                \\ 
        $\sigma_t$                   & $0.127^{+0.012}_{-0.011}$         & $0.187^{+0.009}_{-0.013}$         & $0.187^{+0.009}_{-0.016}$          \\ 
        \hline
        $m_{\log Z}$                 & $-0.069\pm0.008$                  & $-0.098^{+0.004}_{-0.002}$        & $-0.023\pm0.010$                   \\ 
        $Z_{\mathrm{old},0}$         & $0.96\pm0.03$                     & $1.31\pm0.04$                     & $1.19\pm0.05$                      \\ 
        $\sigma_{\log Z}$            & $0.040^{+0.005}_{-0.006}$         & $0.049^{+0.001}_{-0.002}$         & $0.047^{+0.002}_{-0.005}$          \\ 
        \hline
        $A_{V,0}$                    & $1.13\pm0.03$                     & $0.17\pm0.04$                     & $1.30\pm0.05$                      \\ 
        $A_{V,\mathrm{max}}$         & $0.27\pm0.02$                     & $0.27\pm0.05$                     & $0.35\pm0.07$                      \\ 
        $\sigma_{A_V}$               & $0.217\pm0.008$                   & $0.256^{+0.014}_{-0.013}$         & $0.286^{+0.017}_{-0.016}$          \\ 
        \hline
        $m_{\log \sigma}$            & $-0.097\pm0.002$                  & $-0.097^{+0.003}_{-0.002}$        & $-0.031^{+0.001}_{-0.002}$         \\ 
        $\sigma_{\mathrm{disp},0}$   & $86.0^{+0.5}_{-0.6}$              & $125.8^{+1.7}_{-2.0}$             & $151.1\pm0.9$                      \\ 
        $\sigma_{\log \sigma}$       & $0.011\pm0.001$                   & $0.050\pm0.003$                   & $0.009^{+0.002}_{-0.001}$          \\ 
        \hline
    \end{tabular}
    \label{tab:hyper_results}
\end{table}
\renewcommand{\arraystretch}{1.0}

\section{Results} \label{sec:results}
We fit the Voronoi binned MaNGA spaxels of the three PSBs 7965-1902, 12067-3701 and 12514-3702 using the hierarchical model described in Section \ref{sec:Hbayes_methods}. The fitted population properties for all hyper-parameters are reported in Table \ref{tab:hyper_results}. 

\begin{figure*}
    \centering
    \includegraphics[width=0.90\textwidth]{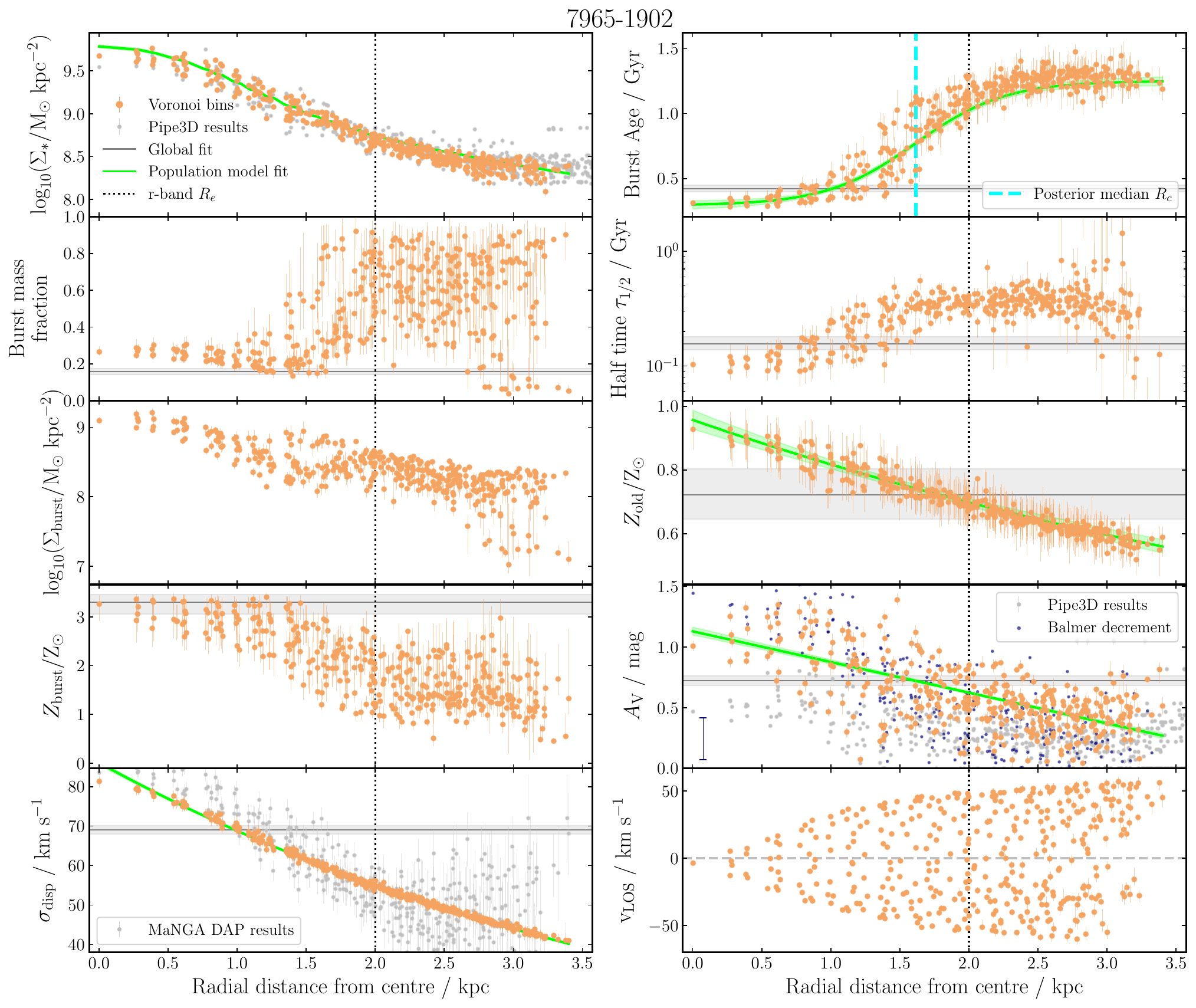}
    \includegraphics[width=\textwidth]{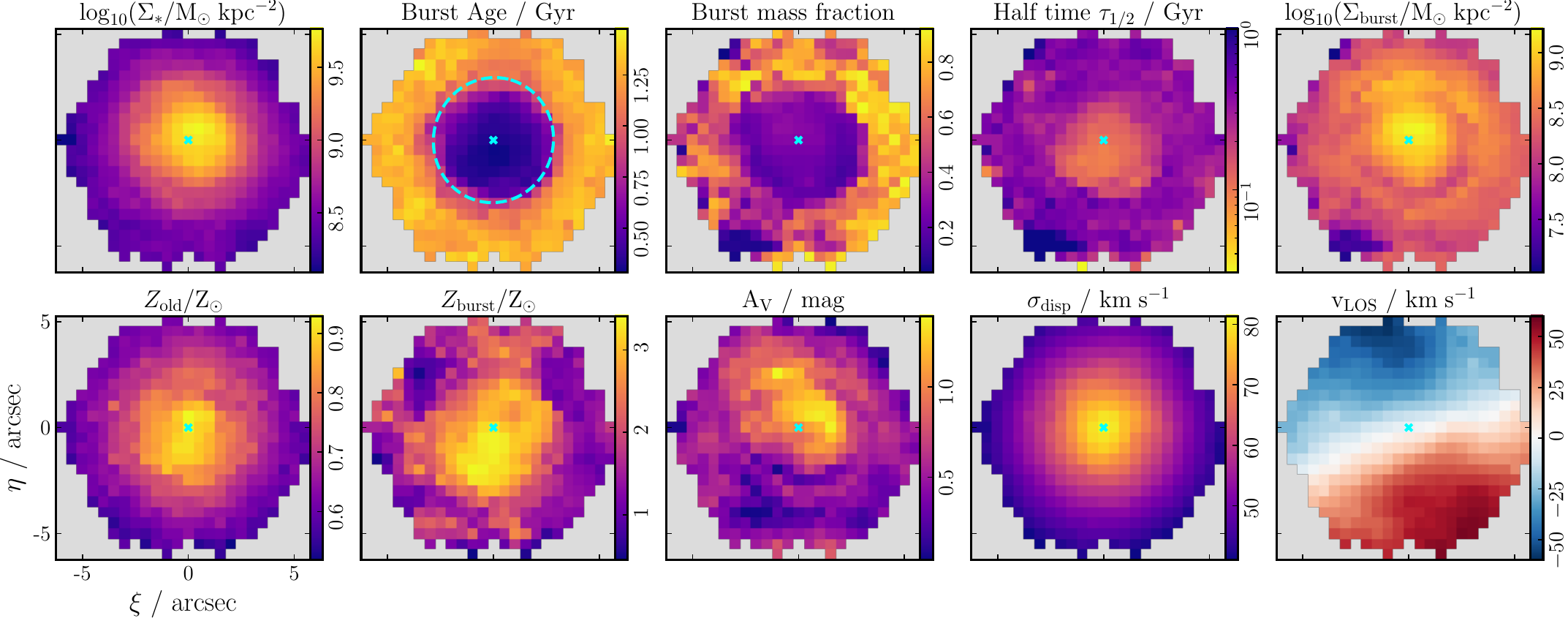}
    \caption{Resolved properties of 7965-1902 using the hierarchical model. \textbf{Top}: Radial gradients of 10 selected properties (see main text for their description). In all panels, the posterior median and $1\sigma$ uncertainty of the Voronoi bins are shown as orange dots and error bars. For model-independent properties, measurements of all spaxels individually from pipe3D are shown as light grey dots (see main text for caveats on $A_V$). $A_{V,\mathrm{ISM}}$ measured via the Balmer decrement and assuming the MAP $\eta$ value from the posterior of the global fit is shown as dark blue dots, with the arithmetic mean of the $1\sigma$ uncertainty shown in the bottom left of the $A_V$ panel. As in Fig. \ref{fig:justify_tburst_gradient}, results from the global fit are shown as a horizontal grey line and shaded band. For the five parameters that we assigned population models to, the population model's posterior median and $1\sigma$ uncertainty are shown as a lime line and shaded region. The galaxy's $r$-band 2D S\'ersic $R_e$ from the NSA catalogue is shown as a black vertical dotted line. In the top right, a cyan vertical dashed line marks the population model's posterior median $R_c$ estimate ($t_\mathrm{burst}$ radial profile inflection point). \textbf{Bottom}: Maps of the posterior median estimates for the same properties. Compared to the others, in the half time panel the direction of the colour map is reversed and in $\log_{10}$ scaling. Cyan crosses mark the central spaxel (the spaxel with the lowest luminosity-weighted elliptical polar distance from the galaxy's centre), while the cyan ellipse marks the posterior median $R_c$ value.}
    \label{fig:results_7965}
\end{figure*}

\begin{figure*}
    \centering
    \includegraphics[width=0.92\textwidth]{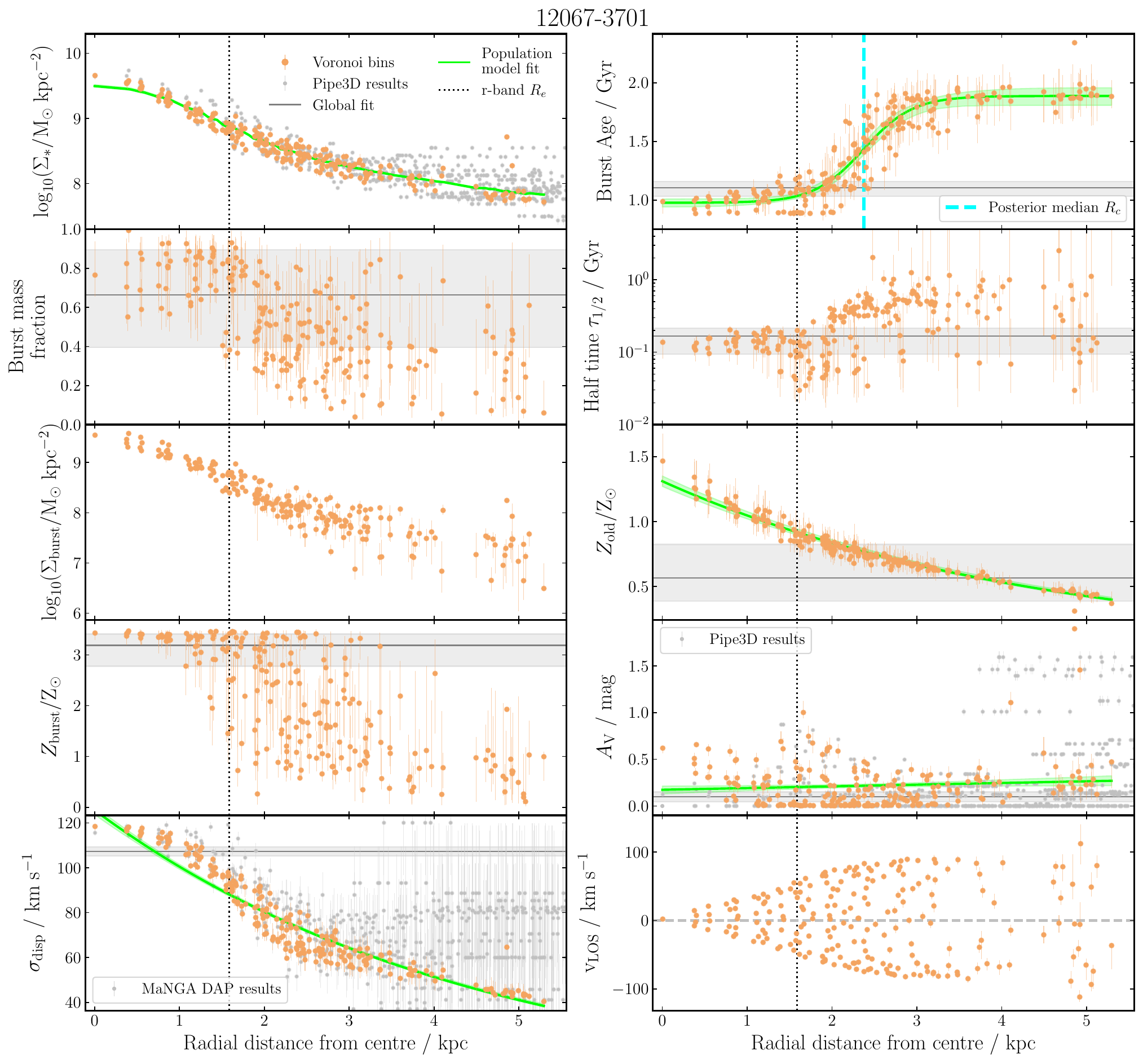}
    \includegraphics[width=\textwidth]{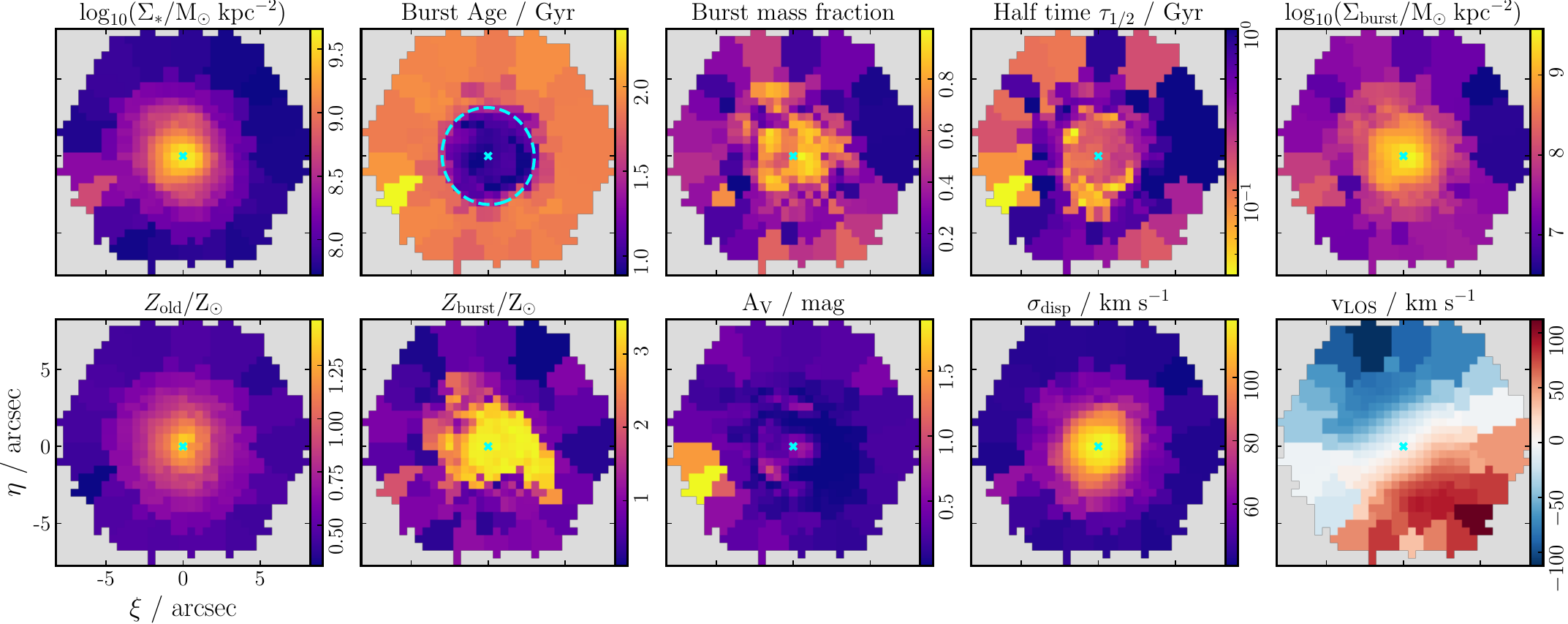}
    \caption{Resolved properties of 12067-3701 using the hierarchical model, in the same style as Fig. \ref{fig:results_7965}. Note that $A_V$ values measured from Balmer decrement are not shown for this galaxy due to low H$\beta$ SNR ($<5$) in all spaxels.}
    \label{fig:results_12067}
\end{figure*}

\begin{figure*}
    \centering
    \includegraphics[width=0.92\textwidth]{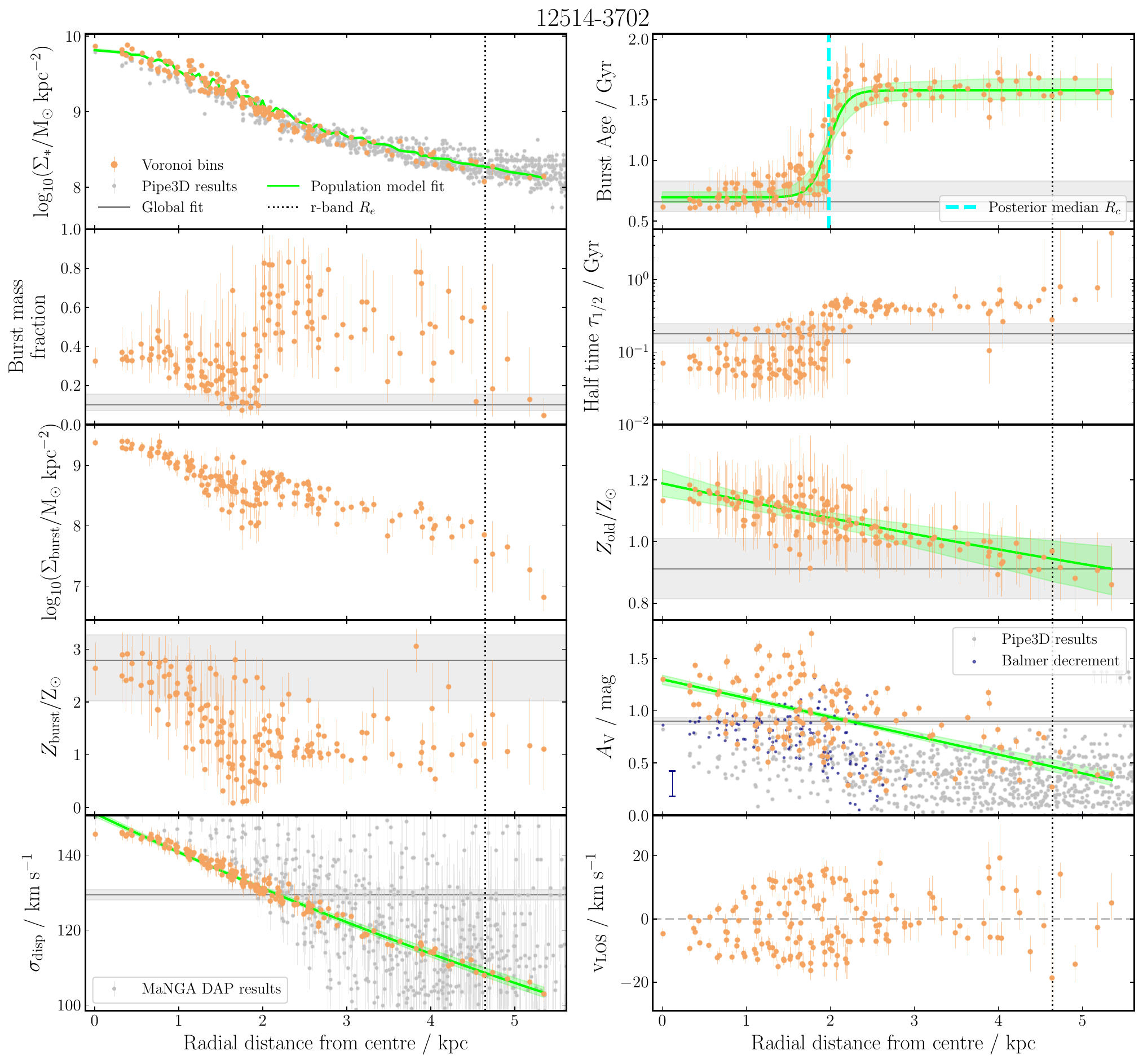}
    \includegraphics[width=\textwidth]{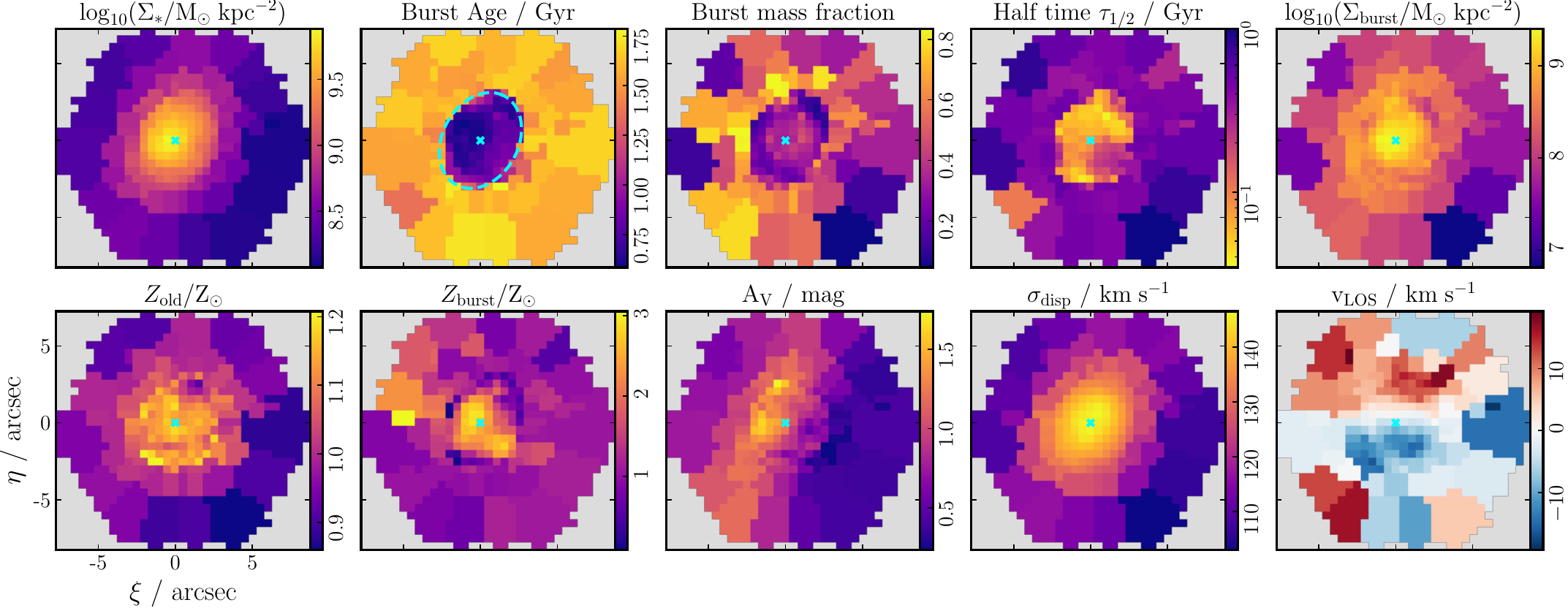}
    \caption{Resolved properties of 12514-3702 using the hierarchical model, in the same style as Fig. \ref{fig:results_7965}.}
    \label{fig:results_12514}
\end{figure*}

We show the fitted radial profiles of 10 local properties in the top portions of Figs. \ref{fig:results_7965}, \ref{fig:results_12067} and \ref{fig:results_12514}. From top to bottom and left to right are the radial gradients of the stellar mass surface density\footnote{Note that here we show the surface density of the total stellar mass formed in each region, rather than the more usual light profile, or the surface density of the living stellar mass.} ($\Sigma_*$), burst age ($t_\mathrm{burst}$), burst mass fraction ($f_\mathrm{burst}$), quenching timescale measured by the duration it takes for the SFR to drop from the starburst's peak value to half of that SFR ($\tau_{1/2}$), surface density of the burst mass ($\Sigma_\mathrm{burst}$), pre-burst metallicity ($Z_\mathrm{old}$), post-burst metallicity ($Z_\mathrm{burst}$), ISM dust extinction in $V$-band ($A_V$), velocity dispersion ($\sigma_\mathrm{disp}$) and stellar velocity along the line of sight ($\mathrm{v_{LOS}}$).
We again stress that other than for $\Sigma_*$, we did not directly model the effects of PSF and the radial profiles shown are PSF-convolved.

We show maps of the same 10 fitted resolved properties in the lower sections in the same figures. Maps of the their $1\sigma$ uncertainties are provided in Appendix \ref{apx:uncertainty_maps}.

We emphasize that because of the usage of BHM when fitting the resolved spectra, we have successfully extracted population-level information concerning the radial profiles of various resolved galaxy properties, simultaneously allowing non-axisymmetric patterns to be uncovered as seen in some of the map panels in Figs. \ref{fig:results_7965}, \ref{fig:results_12067} and \ref{fig:results_12514}. The fitting of the full datacube under a joint coherent framework also improved the estimation precision of individual Voronoi bins' parameters, particularly those at the outskirts with lower SNR.

\subsection{Structural properties} \label{sec:structural_properties}
As shown in the top left panels of Figs. \ref{fig:results_7965}, \ref{fig:results_12067} and \ref{fig:results_12514}, all three galaxies have a negative gradient in stellar mass surface density as expected. 
We measured S\'ersic indices between 3.5 and 5.5 from the three galaxies (see Table \ref{tab:hyper_results}). These values are higher than the $n\simeq1$ value of most disc galaxies measured from light profiles, and are instead more consistent with lenticular and elliptical galaxies \citep{nair2010,fischer2019}. This agrees with earlier studies that measured Sersic indices of PSBs from SDSS images \citep{quintero2004,mendel2013,pawlik2018} and Hubble images \citep{yang2004,yang2008}, and those from earlier PSBs at $z\sim0.7$ \citep{setton2022}. Therefore, similar to other PSBs, our three PSBs have already experienced a morphological transformation from late-type to early-type galaxies.

From our stellar mass surface density modelling, for two galaxies (7965-1902 and 12067-3701) we measure half-mass radii that are larger than their half-light radii obtained from their SDSS $r$-band images, while our measured half-mass radius of the third galaxy, 12514-3702, is comparable to its half-light radius within $1\sigma$ uncertainty (comparing Table \ref{tab:data} to Table \ref{tab:hyper_results}). The larger half-mass radii of the first two galaxies implies they exhibit positive mass-to-light ratio gradients, which is likely a result of their positive stellar age gradients (Section \ref{sec:results2}). However, our estimated half-mass radii of these two galaxies lie beyond the radial distance covered by the MaNGA observations. Therefore, they completely rely on the extrapolation of the stellar mass surface density profile within the radial range covered outwards assuming the S\'ersic profile.

We compare our measured stellar mass surface densities to those derived by a different SED fitting code, pipe3D \citep{manga-pipe3d1,manga-pipe3d2}, which are plotted in the top sections of Figs. \ref{fig:results_7965}, \ref{fig:results_12067} and \ref{fig:results_12514} as smaller gray dots. Good agreement is reached between the two methods' $\Sigma_*$ estimates. 

Examining the maps for the line-of-sight velocities ($\mathrm{v_{LOS}}$) in the lower sections of Figs. \ref{fig:results_7965}, \ref{fig:results_12067} and \ref{fig:results_12514}, 7965-1902 and 12067-3701 show stellar kinematics that are consistent with rotating discs. The $\mathrm{v_{LOS}}$ map of 12514-3702 shows a more disturbed rotation, particularly in the outskirts. This galaxy has the lowest axis ratio b/a among the three as measured in the NSA catalogue, suggesting that its weaker rotation is not likely to be an inclination effect. 
We measure the baryon projected specific angular momentum ($\lambda_R$) within $R<R_e$ ($r$-band $R_e$) following the methods in \cite{emsellem2007}. For 7965-1902 and 12067-3701, we measure $\lambda_{R_e}\sim0.34$ and $\lambda_{R_e}\sim0.19$, respectively, indicating fast rotation ($\lambda_{R_e}>0.1$). For 12514-3702, we measure $\lambda_{R_e}\sim0.06$, indicating slow rotation ($\lambda_{R_e}<0.1$). 
Therefore, 2/3 of our PSBs show strong stellar rotation, which is consistent with early-type galaxies where a majority of them retain clear stellar rotation \citep[e.g.][]{krajnovic2011,emsellem2011}. The most disturbed galaxy, 12514-3702, displays likely tidal features \citep{vazquez-mata2025}, consistent with the disturbed rotation being related to recent galaxy interactions.

In the velocity dispersion panels (top section, bottom left) of Figs. \ref{fig:results_7965}, \ref{fig:results_12067} and \ref{fig:results_12514}, we observe that all three PSBs follow a tight negative gradient in velocity dispersion. The gradient's negative nature agrees with the average radial profile of early-type galaxies \citep{falcon-barroso2017}, and could be consistent with $\log_{10}(M_*/\mathrm{M_\odot})>10$ late-type galaxies (\citealt{parikh2021}, but see also \citealt{falcon-barroso2017}). 
12514-3702 has the highest central velocity dispersion among the three (Table \ref{tab:hyper_results}), and is also the most massive (Table \ref{tab:data}). This is consistent with literature findings where more massive galaxies have higher central velocity dispersion \citep[e.g.][]{mogotsi2019,parikh2021}.

Our estimated $\sigma_\mathrm{disp}$ are in general comparable to the results from MaNGA DAP of individual spaxels measured through \texttt{pPXF} \citep{ppxf1,ppxf2}. Disagreements can be found in the central regions of 7965-1902 and 12514-3702, where our estimates are $>2\sigma$ lower than those from MaNGA DAP. Some very massive ($\log_{10}(M_*/\mathrm{M_\odot})>11$) early-type galaxies have been found to display U-shaped stellar velocity dispersion gradients, where at large radii the negative gradient flattens before turning positive \citep{veale2018}. The increased upward scatter of the MaNGA DAP velocity dispersions at $R>2.5\;$kpc in 7965-1902 and 12067-3701 could be a hint of this upturn, but this is subject to very large estimation uncertainties. 

\subsection{An outside-in starburst and quenching sequence} \label{sec:results2}
We find all three galaxies to have experienced a prominent, recent central starburst. The central burst ages are estimated as $\sim0.3\,$Gyr, $\sim1.0\,$Gyr and $\sim0.7\,$Gyr for 7965-1902, 12067-3701 and 12514-3792, respectively. Figs. \ref{fig:results_7965}, \ref{fig:results_12067} and \ref{fig:results_12514} shows that the central burst mass fractions vary greatly between the galaxies, from $\sim0.3$ in 7965-1902 and 12514-3792 to $>0.6$ in 12067-3701. Following the starburst, they all experienced rapid quenching, with the estimated central quenching timescale as $\tau_{1/2}\simeq100\,$Myr. 

The outer regions of all three galaxies also experienced an increase in star formation before declining in SFRs. Figs. \ref{fig:results_7965}, \ref{fig:results_12067} and \ref{fig:results_12514} shows that this outer starburst occurred $\sim1\,$Gyr before the central starburst in all three galaxies, but experienced comparatively slower quenching ($\tau_{1/2}\geq300$).

In 7965-1902 and 12514-3792, the starburst in the central regions contributed a lower stellar mass fraction ($f_\mathrm{burst}$) than those in the outer regions, while the opposite is observed in 12067-3701. However, due to the higher stellar mass surface density in the centre of all three galaxies, all central regions formed higher total stellar masses during their starbursts compared to their outer regions (surface density of burst mass panel in Figs. \ref{fig:results_7965}, \ref{fig:results_12067} and \ref{fig:results_12514}).

As shown in the top sections of Figs. \ref{fig:results_7965}, \ref{fig:results_12067} and \ref{fig:results_12514}, we observe discontinuities in the radial trends of some stellar properties, particularly strongly in 12514-3702 (burst mass fraction in 7965-1902; $\tau_{1/2}$ in 12067-3701; burst age, burst mass fraction and $\tau_{1/2}$ in 12514-3702). Therefore, we consider the central and outer starbursts in all three galaxies to be distinct episodes of increases in star formation (see also Fig. \ref{fig:justify_tburst_gradient}). We mark the elliptical polar distance from each galaxy's centre that corresponds to the posterior median inflection point radius ($R_c$) of our burst age radial model in Figs. \ref{fig:results_7965}, \ref{fig:results_12067} and \ref{fig:results_12514}. Regions inside the ellipses are dominated by an episode of more recent starburst, while those outside the ellipses are dominated by an earlier rise in star formation. This is consistent with the single-fibre results from \cite{french2018}, which found that a significant fraction of local PSBs were better fit with a SFH having two starbursts. The starburst patterns observed could arise from one continuous process or two events affecting the outskirts and the centre at different times. The common burst age gap between the three galaxies suggests that the starburst and subsequent quenching in all three galaxies could be driven by a similar mechanism. 

Comparing the gray bands to the orange dots in the burst age, burst mass fraction and half time decay panels in the top sections of Figs. \ref{fig:results_7965}, \ref{fig:results_12067} and \ref{fig:results_12514}, it is clear that the estimates from the global fit follow the properties of the central regions but differ from the outer regions. Thus, the outer, older starbursts are completely undetected by the global fit, which measures a SFH that closely resembles that of the central regions. This central bias is due to the higher central stellar mass surface density and the younger starburst outshining the older peripheral burst (see also Fig. \ref{fig:justify_tburst_gradient}). This result highlights that when fitting the spatially-integrated SED of a galaxy, not just spatial patterns, but also temporal patterns in the galaxy's properties could be completely missed (see also \citealt{mosleh2025}).

In all three galaxies, compared to the central, younger burst regions, the outer older burst regions are found to have generally longer quenching timescales. This could suggest that the central regions were quenched by rapid physical mechanisms, whereas the outer regions were subject to slower processes. Alternatively, unlike the centre, the outer regions could have maintained a low level of star formation, perhaps driven either by weak gas inflow or incomplete quenching.

In Fig. \ref{fig:results_12514}, the radial profiles of many local properties in 12514-3702 show a strong discontinuity at $R\simeq2\;$kpc (burst age, burst mass fraction, half time, $Z_\mathrm{burst}$). This discontinuity is larger than those seen in the other two galaxies. This difference between the galaxies is unlikely to be an inclination effect: if this discontinuity exists in all three galaxies, given that 12514-3702 has the lowest axis ratio b/a among the three as measured in the NSA catalogue, this galaxy should show the weakest discontinuity. Therefore, the strong discontinuity is a unique property of 12514-3702. Interestingly, this is the galaxy that is morphologically identified by \citet{vazquez-mata2025} as having tidal features. Its burst age and burst mass fraction are intermediate between the other two galaxies, so this is not simply interpreted as due to a more gas rich merger, or having had less time to undergo less mixing of stars within the disk. Perhaps this indicates a merger orientation that is more conducive to causing long-lasting tidal features. A larger sample would be required to see how common this feature is, and comparison to simulations may help to tease out the likely cause. 

As seen in the burst age maps in the lower sections of Figs. \ref{fig:results_7965}, \ref{fig:results_12067} and \ref{fig:results_12514}, the burst age radial gradient is generally symmetric. On the other hand, the burst mass fraction is relatively symmetric in the region with the recent central starburst, but highly asymmetric in the outer regions of all three galaxies. For example, arcs that span $\sim180^{\circ}$ with higher $f_\mathrm{burst}$ than neighbouring regions can be observed in 7965-1902 and 12514-3702. The south-east side of 12067-3701 (bottom left in Fig. \ref{fig:results_12067}) has higher burst mass fractions and stellar mass surface density than its north-west regions. These features resemble spiral arms or tidal features, where higher SFRs could have taken place \citep[e.g.][]{roberts1969,schombert1990,walter2006}.

\subsection{Resolved evolution in stellar metallicity}
In the pre-burst metallicity panels in each of the top sections of Figs. \ref{fig:results_7965} to \ref{fig:results_12514}, all three PSBs exhibit a distinctly negative gradient in stellar metallicity prior to the starburst. Their centres were solar or slightly super-solar, while their outskirts were sub-solar. The negative gradient is consistent with most late-type galaxies measured from IFS observations \citep{sanchez-bazguez2014,gonzalez-delgado2015,parikh2021}. Hence, our three PSBs are consistent with originating from typical disc galaxies prior to the starburst and quenching events.

Generally, negative gradients are observed in the post-burst metallicity of all three galaxies, but with increased scatter compared to the pre-burst metallicities, possibly driven by the large uncertainties in the outskirts. All central regions have higher post-burst metallicity than their outer regions, reaching super-solar levels while the outskirts are around solar metallicities. 

Comparing pre- and post-burst metallicities, the central regions experienced a stronger metallicity increase (7965-1902: $\sim+2\mathrm{Z_\odot}$, 12067-3701: $\sim+1.75\mathrm{Z_\odot}$, 12514-3702: $\sim+2\mathrm{Z_\odot}$), while the outer regions experienced weaker increases or remained constant (7965-1902: $\sim+1.25\mathrm{Z_\odot}$, 12067-3701: $\sim+0.5\mathrm{Z_\odot}$, 12514-3702: $\sim+0$). This suggests that the later, central starbursts are connected with significantly stronger enrichment of the local ISM than the earlier, outer starbursts.

\subsection{Dust properties}
In the $A_V$ (ISM dust attenuation) panels in each of the top sections of Figs. \ref{fig:results_7965}, \ref{fig:results_12067} and \ref{fig:results_12514}, we found large scatter in the dust attenuation strength in all three galaxies. 7965-1902 and 12514-3702 display negative gradients, while 12067-3701 shows a flat gradient. Negative dust attenuation gradients have been reported in observations of star-forming galaxies \citep{nelson2016,liu2017,greener2020}.

In the $A_V$ maps of 7965-1902 and 12067-3701 (Figs. \ref{fig:results_7965} and \ref{fig:results_12067}), complex non-axisymmetric dust features including shells and lobes of increased $A_V$ can be observed. In 12514-3702, the bottom middle panel of Fig. \ref{fig:results_12514} shows that this galaxy likely has a prominent north-south dust lane that is not visible in the SDSS image. These non-axisymmetric features contribute to the large scatter in the galaxies' $A_V$ radial profiles mentioned above. This complex dust geometry could potentially lead to an underestimation in the SFRs of PSBs when estimated from integrated spectroscopy \citep{valeasari2020}.

We observe mild correlations between the galaxies' dust distributions with the stellar properties of the burst. In the lower section of Fig. \ref{fig:results_7965}, regions of high $A_V$ in 7965-1902 mostly overlap with regions of low burst age, low burst mass fraction and high post-burst metallicity. In the lower section of Fig. \ref{fig:results_12067}, the central lobes of high $A_V$ in 12067-3701 overlap with some of the lobes of high burst mass fraction. In the literature, studies generally find dust attenuation strength to increase with metallicity \citep{theios2019,maheson2024}. Therefore, the high post-burst metallicities in the centres of the three galaxies and the high $A_V$ in some regions could be caused by the same process.

Our $A_V$ estimates in 7965-1902 and 12514-3702 are typically higher than those from pipe3D (top sections, Fig. \ref{fig:results_7965} and \ref{fig:results_12514}), likely due to different assumptions in the dust law and dust attenuation from birth clouds. To investigate this, we obtained independent estimates of ISM $A_V$ through the Balmer decrement method \citep{kennicutt1992} for spaxels with H$\beta$ SNR $>5$. First, we obtain the H$\alpha$ and H$\beta$ emission line fluxes for all spaxels from the MaNGA DAP. Then, assuming Case B recombination ($F_\mathrm{H\alpha}/F_\mathrm{H\beta}=2.87$) and the same \cite{wild2007} dust law as our spectral fitting, we calculate the birthcloud $A_V$. Finally, we convert the total attenuation to ISM $A_V$ by setting $\eta$ as the MAP value from the global fits. These values are marked with dark blue dots in Figs. \ref{fig:results_7965} and \ref{fig:results_12514}. No spaxels in the observation of 12067-3701 have H$\beta$ SNR $>5$, therefore we do not show any Balmer decrement $A_V$ in Fig. \ref{fig:results_12067}. From Figs. \ref{fig:results_7965} and \ref{fig:results_12514}, the $A_V$ estimates from Balmer decrement are more consistent with our $A_V$ estimates, which suggests that the BHM approach is beneficial for measuring these more tricky regional properties. 

\section{Discussion} \label{sec:discussion}
In this section, we first briefly showcase the advantages of our BHM approach over traditional methods. Then, we explore the implications of our results for the commonly used spatial descriptors ``inside-out'' and ``outside-in'' quenching, in order to frame our results in language common in the field. Finally, we compare the spatially resolved star formation history patterns observed in the three PSBs to previous observations of merger and post-merger galaxies, and simulation predictions.

\subsection{Comparing the hierarchical approach to non-hierarchical fitting} \label{sec:discussion_hbayes}
\begin{figure*}
    \centering
    \includegraphics[width=\textwidth]{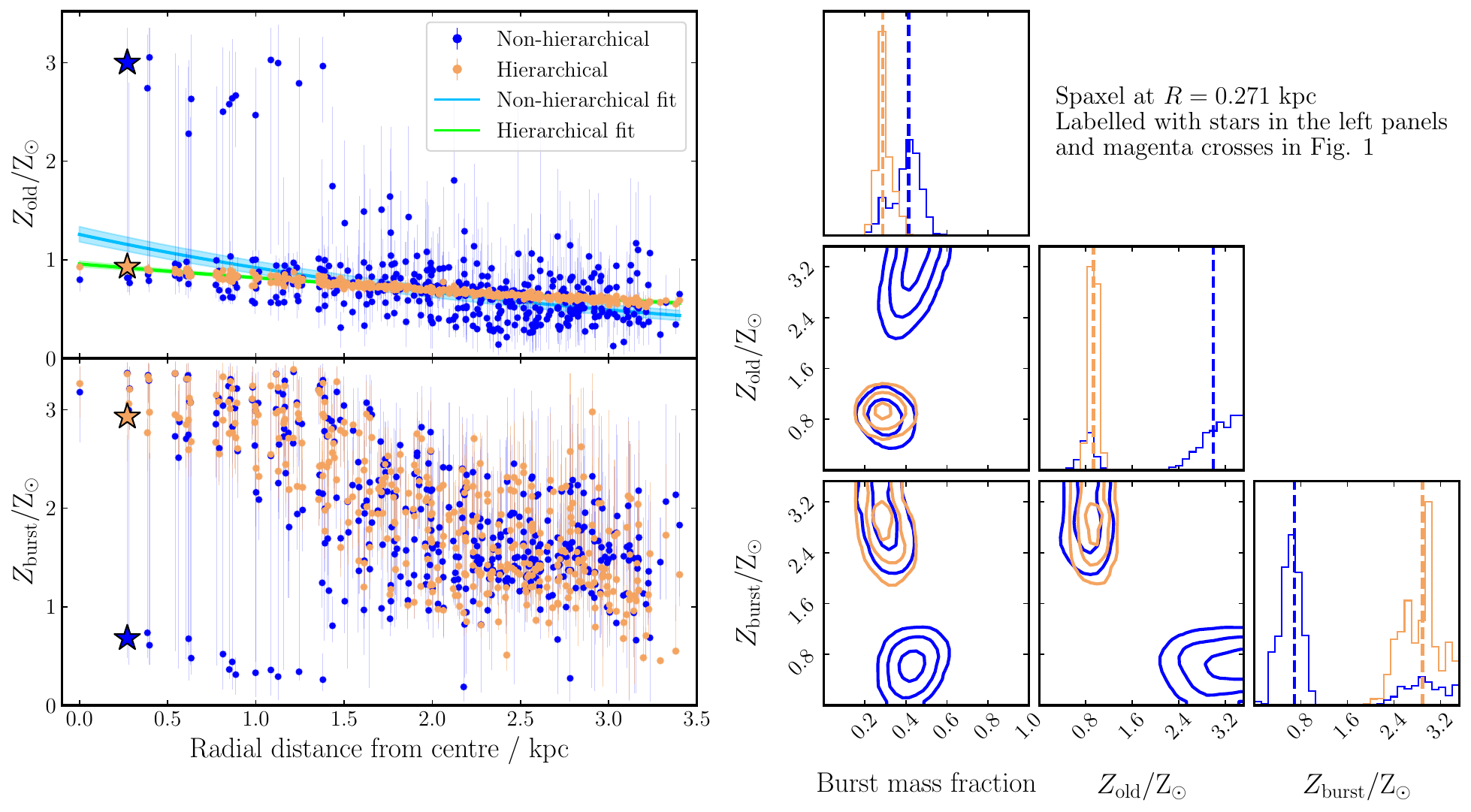}
    \caption{Comparing results from fitting all Voronoi bins of 7965-1902 independently (non-hierarchical, blue) and jointly under the hierarchical model (orange). \textbf{Left}: Radial gradients in pre-burst metallicity (top) and post-burst metallicity (bottom). The light blue curve shows a least squares fit to the posterior median values of the independent pre-burst metallicity estimates. The blue shaded region marks this fit's $1\sigma$ region. The lime curve marks the hierarchical model's fit (posterior median). The large stars mark the locations of the Voroni bin shown in the right. \textbf{Right}: 2D posterior density distributions of a spaxel at $R=0.271\;$kpc (127\textsuperscript{th} Voronoi bin, location marked with magenta crosses in Fig. \ref{fig:Hbayes_PSBmaps}) for burst mass fraction, pre-burst metallicity and post-burst metallicity. In the diagonal panels, the histograms show the 1D marginalised posterior distributions, and the vertical dashed lines mark the distributions' medians. In the off-diagonal panels, the contour lines correspond to (1,2,3)$\sigma$ regions (enclosing the top 39.3, 86.4, and 98.9 per cent of marginalised posterior probability).}
    \label{fig:Bayes_vs_HBayes}
\end{figure*}

Before addressing the implications of our results, we first briefly showcase the advantages of the BHM approach over modelling the Voronoi bins from each galaxy independently. Fig. \ref{fig:Bayes_vs_HBayes} compares the results for 7965-1902 from \textit{stage 1}, which represents the non-hierarchical approach (blue in all panels), to the results from the full BHM method (orange in all panels). We stress that both approaches were provided with the same data and \textsc{Bagpipes} modelling configuration. 

The first advantage is the simultaneous modelling of both local and population properties, which ensures that, in principle, uncertainties at the local Voronoi bin level are properly propagated through to the population parameters. To illustrate this, in the top left panel of Fig. \ref{fig:Bayes_vs_HBayes}, we perform a least squares fit to the posterior median estimates from the non-hierarchical approach (light blue curve and shaded region), assuming the equation used to model the $Z_\mathrm{old}$ radial gradient in the BHM approach (Equation \ref{eq:zmet_old_gradient}). This curve deviates from the radial relationship we measured using the BHM approach (lime curve). The difference appears to be significant at low radial distances (disagreement $>2\sigma$), because the uncertainties of the measurements at each Voronoi bin were not properly propagated through to this fit. This issue is overcome when using the BHM approach.

Secondly, the assertion that properties in local regions are mutually correlated through the population model allows degeneracies in the local regions to be broken. In the right panels of Fig. \ref{fig:Bayes_vs_HBayes}, we compare the 2D posterior density distributions of a spaxel at $R=0.271\;$kpc (marked in the left panels using large stars, and the position is marked with magenta crosses in Fig. \ref{fig:Hbayes_PSBmaps}) from both fitting approaches. The two loci of high-probability regions in the 2D posterior density distributions of the non-hierarchical approach, and their bimodal distributions in the 1D posterior distributions of $Z_\mathrm{old}$ and $Z_\mathrm{burst}$ suggest that the data is consistent with two distinct solutions. One has a high $Z_\mathrm{old}$ and a low $Z_\mathrm{burst}$, while the other has the opposite. These two solutions are degenerate. In contrast, in the BHM approach, this degeneracy is broken since only one of the solutions, low $Z_\mathrm{old}$ and high $Z_\mathrm{burst}$, is consistent with the radial pattern observed in all other Voronoi bins.

Thirdly, adopting a BHM approach can significantly lower the uncertainties of the local properties, particularly for Voronoi bins with lower SNR (also seen by \citealt{gonzalez-gaitan2019} with a similar method on IFS data). In Section \ref{sec:structural_properties}, we noted that our measured stellar mass surface densities of the three galaxies are in excellent agreement with the values measured by pipe3D from fitting all spaxels independently. This is an example of a well-constrained property for which the BHM approach confers no additional benefit. However, in the same section, we also noted that our estimates of $\sigma_\mathrm{disp}$ have significantly lower uncertainties than the non-hierarchical fits from MaNGA DAP. While the increased SNR of the spectra through Voronoi binning undoubtedly contributed\footnote{The shrinking of the prior using results from the global fit have also helped when comparing between our BHM results to those from an independent method (e.g. DAP results). However, both the non-hierarchical and hierarchical results in Fig. \ref{fig:Bayes_vs_HBayes} used the updated priors from the global fit, thus we do not mention this in the main text.}, the BHM method also helped in shrinking the uncertainties by pooling together information from all Voronoi bins (``borrowing strength''). A similar effect can be observed in the top left panel of Fig. \ref{fig:Bayes_vs_HBayes}, where the uncertainties of $Z_\mathrm{old}$ at the outer regions are significantly lower in the BHM approach than the non-hierarchical approach. This effect is not limited to parameters that we directly modelled with radial dependence. In the right panels of Fig. \ref{fig:Bayes_vs_HBayes}, the width of the posterior distribution of burst mass fraction from the BHM approach is considerably smaller than that from the non-hierarchical approach, resulting from the degeneracy breaking as discussed above.

We measure the change in posterior 1D distributions between the non-hierarchical and hierarchical fits of the ten parameters shown in Figs. \ref{fig:results_7965}, \ref{fig:results_12067} and \ref{fig:results_12514} using the Jenson-Shannon (JS) distance, repeated for all Voronoi bins and all galaxies. Table \ref{tab:JS_distance} lists the median JS distance values. In all three galaxies, we found $\mathrm{v_{LOS}}$ to have experienced the smallest change in its posterior distributions (lowest median JS distance), while $Z_\mathrm{old}$ experienced the largest change (highest JS distance). This is unsurprising. Typically, $\mathrm{v_{LOS}}$ can be easily constrained with the observed wavelength of key spectral features, leading it to only minimally benefit from the hierarchical model. In contrast, $Z_\mathrm{old}$ only introduces fine metal absorption features on the observed spectrum, thus can be highly uncertain and degenerate with other parameters (Fig. \ref{fig:Bayes_vs_HBayes}). Therefore, it benefits from the advantages of modelling the galaxy as a population as discussed above.

\begin{table}
    \centering
    \caption{Median Jenson-Shannon distance between the non-hierarchical and hierarchical posterior distributions. The median is over all Voronoi bins in each galaxy.}
    \begin{tabular}{llll}
        \hline
        Plate-IFU & 7965-1902 & 12067-3701 & 12514-3702  \\
        \hline
        $\log_{10}(\Sigma_*)$ & 0.38 & 0.32 & 0.31 \\
        Burst Age & 0.44 & 0.47 & 0.36 \\
        Burst mass fraction & 0.37 & 0.31 & 0.29 \\
        $\tau_{1/2}$ & 0.32 & 0.33 & 0.26 \\
        $\log_{10}\Sigma_\mathrm{burst}$ & 0.37 & 0.31 & 0.28 \\
        $Z_\mathrm{old}$ & 0.56 & 0.61 & 0.53 \\
        $Z_\mathrm{burst}$ & 0.35 & 0.30 & 0.29 \\
        $\sigma_\mathrm{disp}$ & 0.52 & 0.35 & 0.39 \\
        $\mathrm{v_{LOS}}$ & 0.23 & 0.21 & 0.17 \\
        \hline
    \end{tabular}
    \label{tab:JS_distance}
\end{table}

Here, it is important to note that the advantages of the BHM approach require the population model(s) assumed to be an adequately correct description of the truth. The apparent benefits of shrinking uncertainties and breaking degeneracies can be obtained even if we assume incorrect radial dependences, such as asserting a constant stellar mass surface density profile. But this will result in grossly incorrect estimates for all local and population properties. Therefore, it is paramount to support population models assumed with sufficient evidence from the literature or the data, and to avoid adopting an overly-complex model when there is insufficient evidence for it.

\begin{figure*}
    \centering
    \includegraphics[width=\textwidth]{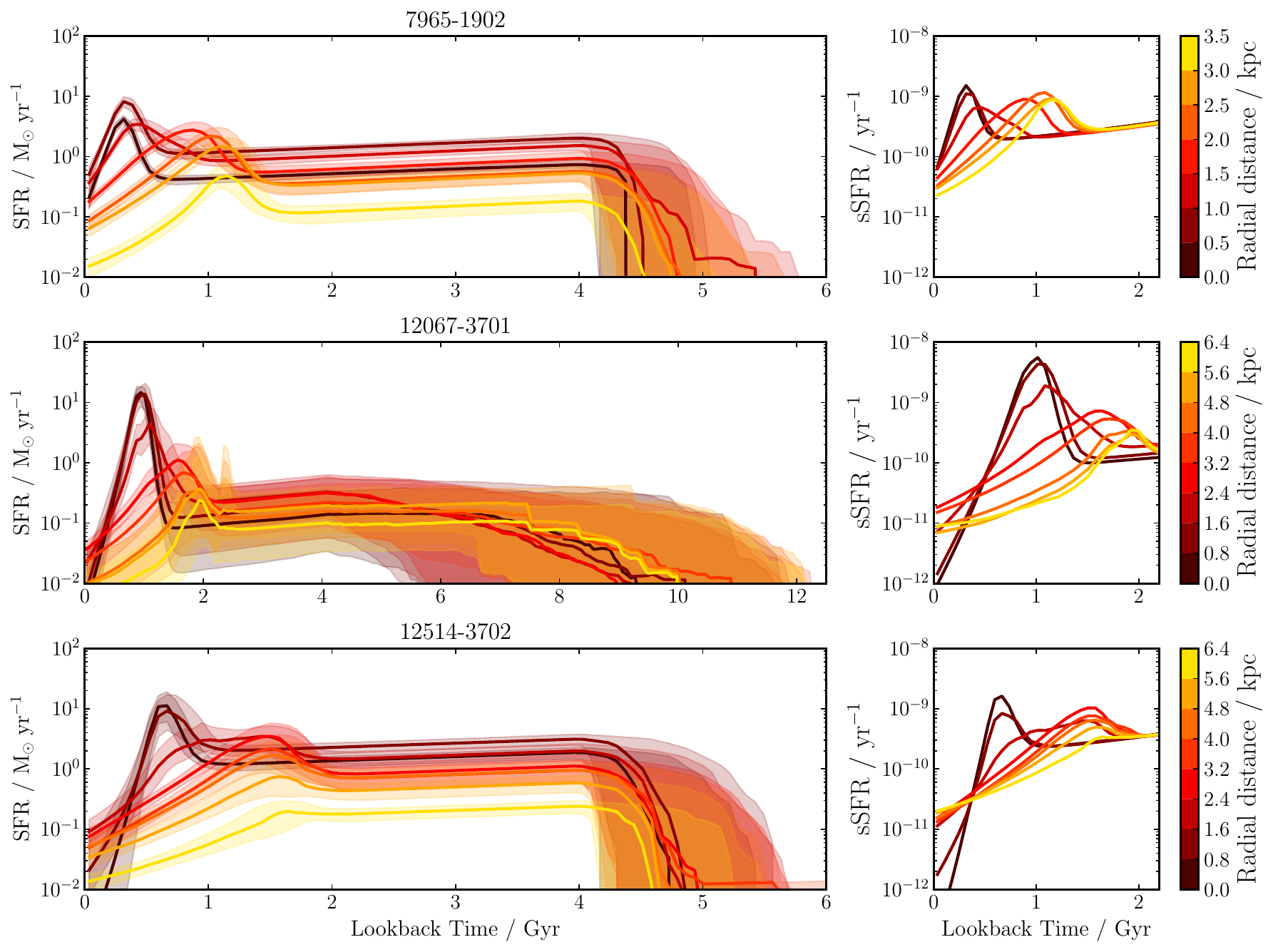}
    \caption{The SFHs of the post-starburst galaxies stacked in bins of increasing radial annuli from the galaxy centre. The \textbf{left} column shows SFR against lookback time. The \textbf{right} column shows sSFR against lookback time, in the most recent $\sim2\;$Gyr. Bins at larger radii are represented with lighter colours. In all panels and radial bins, the posterior median SFH is marked as a solid line, while the $1\sigma$ uncertainty region is marked as a shaded region of the same colour (uncertainty not shown in the right for simplicity). All three galaxies exhibit a two-phase outside-in sequence of starburst and quenching, where the outer regions first undergo starburst that quenches slowly, followed by a stronger central starburst after $\sim1\;$Gyr that quenches more rapidly than the outskirts.}
    \label{fig:radial_sfh}
\end{figure*}

\subsection{Inside-out or outside-in quenching?}
As discussed in Section \ref{sec:intro}, many studies approach the topic of spatially-resolved quenching patterns through the inside-out and outside-in radial trends. Here, we utilise our spatially resolved SFHs to investigate whether the distinction between inside-out and outside-in quenching is useful. 

To aid the interpretation of the results and comparisons to other works, in Fig. \ref{fig:radial_sfh} we bin the fitted SFHs into consecutive radial annuli. In all three galaxies, the outside-in starburst sequence discussed in Section \ref{sec:results2} can be clearly observed. We note that the SFHs shown here do not account for stellar migration, meaning they are not estimates of the true SFRs at the given radii and lookback time.

In Fig. \ref{fig:radial_sfh}, the starburst consistently propagates inwards, in all three PSBs. The decline in SFR thus begins earlier in the outer regions, before the central areas followed suit. This could be viewed as outside-in quenching. However, the outskirts have longer quenching timescales, thus in 2/3 of the PSBs, the outer regions currently have higher SFRs and sSFRs than the centre (12067-3701 and 12514-3702). The third PSB, 7965-1902, will follow suit in the next $\sim500\;$Myr if the SFRs at all radial bins continue to decline at their current rates. Under this consideration, these galaxies can be viewed as inside-out quenching. In fact, if one uses the relative sSFR measured from different radial bins directly as a proxy for inside-out or outside-in quenching, the bottom right panel of Fig. \ref{fig:radial_sfh} shows that depending on when it is observed, 12514-3702 would be labelled as inside-out quenching (lookback time $\sim1.3\;$Gyr), then as outside-in quenching (lookback time $\sim0.6\;$Gyr), and finally back to inside-out quenching (lookback time $=0$).

Several methods have been developed in the literature to select spatial quenching patterns from IFU observations \citep{lin2019,chen2019,li2023}. For our three galaxies, all methods yield ambiguous classifications between inside-out or outside-in quenching. This is because our three galaxies have predominantly quenched spaxels, which hampers the effectiveness of these methods that rely on the relative positions between quenched and non-quenched spaxels. On the other hand, our method can clearly differentiate the timescales over which quenching is happening as a function of radius.

Therefore, the inside-out and outside-in labels are too simplistic to properly capture the complex spatially-resolved burst and quenching patterns in a consistent manner. We urge future studies to take this into account when interpreting sSFR or SFR gradients and, when possible, measure the full resolved SFHs of galaxies and formulate inferences directly from the SFHs.

As a sanity check, we also radially binned the fitted SFHs from the non-hierarchical results (\textit{stage 1}). Compared to Fig. \ref{fig:radial_sfh}, the non-hierarchical results show more uncertain radial SFHs in all outer regions. However, the general observation of the outside-in starburst sequence is still clearly visible, suggesting it is physical in origin.

At this point, it is worth remembering the limitations of our parametric SFH model, which does not include multiple episodic enhancements in star formation. The radial trends shown in Fig. \ref{fig:radial_sfh} are therefore also consistent with the outer regions of the galaxies experiencing two distinct starbursts,  manifesting as a burst with a slow apparent decline rate. In the future, a two-burst SFH model similar to the one in \cite{french2018} could be used to investigate which model is best supported by the data. Ultimately, a step-wise SFH model would be ideal, but this would require significant effort to build into an appropriate population model with sufficient time sampling in the starburst age range \citep[see e.g.][for an example of a step-wise SFH model used in a galaxy population spectral fitting analysis]{leistedt2016}.

\subsection{Testing for an environmental origin}

The outside-in sequence of starburst and quenching in our PSBs might appear consistent with the effects of environmental quenching. In dense galaxy clusters, frequent tidal interactions with other cluster members and the removal of the cold ISM through ram-pressure stripping can lead to quenching beginning from a galaxy's outskirts \citep[e.g.][]{schaefer2017,finn2018,bluck2020}. Indeed, the origin of some PSBs with outside-in quenching characteristics have been attributed to environmental quenching \citep{werle2022,werle2025}. However, from the GEMA-VAC cluster catalogue (\citealt{argudo-fernandez2015}, version DR17), none of our three PSBs have close neighbours nor are members of any clusters and groups, thus rejecting environmental effects as the driver for their recent changes in star formation.

\subsection{Comparison with observed (post-)merger samples} \label{sec:discussion_merger}

While it is generally believed that most local PSBs not residing in galaxy clusters must arise from gas-rich mergers \citep{zabludoff1996,goto2005,pawlik2016,wilkinson2022}, not all show clear merger signatures in their morphology and structure \citep{pawlik2018, ellison2024}. As discussed in Section \ref{sec:data}, only 12514-3702 has been identified as having tidal features by \citet{vazquez-mata2025}, although close inspection of the image of 12067-3701 indicates some outer asymmetry.  Other signs of mergers can be found through internal dust structures in high spatial resolution observations \citep{sazanova2021}, and stellar chemical enrichment from spectral fitting \citep{paper1,paper2}. In the PSBs studied in this paper we find that the most significant starburst is centrally concentrated which, in the merger scenario, corresponds to coalescence of the nuclei. However, our results also indicate significant extended star formation enhancement $\sim1$\,Gyr prior to this. Here we investigate whether the spatial distribution of past star formation provides support or otherwise for the hypothesis that these are post-merger galaxies, despite their faint or absent morphological features. 

Measurements of spatially resolved star formation during mergers have benefitted greatly from the advent of optical IFU data where recent large samples have allowed statistical analyses of SFR enhancements at different merger stages compared to rigorous control samples. \citet{barreraballesteros2015} studied 103 close galaxy pairs in the CALIFA survey, to find a moderate SFR enhancement in the central regions compared to a control sample.  \citet{pan2019} studied a sample of 205 interacting galaxies and 1350 control galaxies in MaNGA to conclude that first passage causes star formation to be triggered predominantly in the inner regions of the systems, with a slight depression at larger radii. At later stages the enhancement in star formation is much more extended. This predominantly central enhancement, with some enhancement at large radii at coalescence when compared to control samples, agreed with \citet{ellison2013} who used central fibre vs. extended photometrically derived SFR to study a large pre-IFU sample of merging and control galaxies. \citet{thorp2019} compared the spatial extent of 36 recent post-mergers (single galaxies visually identified to have tidal tails and shells) in MaNGA, again finding centrally concentrated enhancements in star formation. In this case, some galaxies showed SFR enhancements at large radii, while others showed a depression.
Overall, these results appear to differ from ours, where the SFR enhancement at large radii occurs early on in the merger. 

The enhanced star formation at first passage found in the PSBs is, however, entirely consistent with the results of \citet{cortijoferrero2017c} who studied the spatially resolved star formation histories of four local mergers, including three luminous infrared galaxies (LIRG) and one non-LIRG (the Mice, also presented in \citealt{wild2014}). For the three LIRGs, but not the non-LIRG, they found spatially extended star formation enhancement triggered by first peri-centre passage, in agreement with our results. The outer enhancement factors of 3-7 are also very consistent with the early outer starburst in the PSBs studied here (Fig. \ref{fig:radial_sfh}). As also discussed in \citet{pan2019} the difference between the large-sample results  \citep{ellison2013, barreraballesteros2015, thorp2019, pan2019}, and those of less common high gas fraction LIRGs in \citet{cortijoferrero2017c}, suggests that different galaxy structures, gas fractions, or orbital dynamics of galaxy mergers lead to different spatial and temporal patterns in star formation enhancement. The coincidence of our results with those of the LIRGs in \citet{cortijoferrero2017c} perhaps implies that PSBs selected via strong H$\delta$ absorption and weak H$\alpha$ emission across most of the galaxy as in this paper, are descended from the merger of high gas fraction disk galaxies, which are detected as LIRGs or ULIRGs during their merger.

In \citet{wild2016} and \citet{wild2020} it was argued that $z\simeq1$ PSBs had number densities, stellar masses and recent peak SFRs consistent with being the descendants of sub-mm galaxies.  Similarly, \citet{wilkinson2021} showed that the clustering of $0.5<z<2$ PSBs was consistent with them being hosted by dark matter haloes with the same mass as sub-mm galaxies. Thus, it is perhaps not surprising that there is a correspondence between the spatially resolved SFR properties of $z\simeq0$ PSBs and LIRGs.  \citet{magnelli2013} give number densities of $\sim 10^{-6}\;{\rm Mpc}^{-3}$ for ULIRGs and  $\sim 10^{-4}\;{\rm Mpc}^{-3}$ for LIRGS at $z\simeq0.1$ in the deepest Herschel-PACS survey. This compares to number densities of $\sim 10^{-5.6}\;{\rm Mpc}^{-3}$ for PSBs with stellar masses $>10^{10.6}\;{\rm M_\odot}$ at similar redshifts, although with a different selection method to that used here \citep{rowlands2018a}. While certainly comparable at the order-of-magnitude level, further work would be required to robustly compare these numbers, using comparable stellar mass measurements, and taking into account information about the visibility times of the two samples.

The presence of dust lanes has been shown to be related to recent mergers \citep{goudfrooij1998,shabala2012,kaviraj2013,dariush2016}. We observe complex non-axisymmetric dust geometry in all of our PSBs, again supporting their origin as gas-rich mergers. Additionally, the distribution of burst mass fraction of the central recent starburst in 12607-3701 shows several distinct ``clumps'', consistent with the galaxy's high clumpiness value \citep{vazquez-mata2025}, which could also be a sign of a recent major merger \citep{lotz2004,dimatteo2008,puech2010,calabro2019}.

Overall, we conclude that the PSBs studied here are consistent with arising from major mergers of gas-rich disk galaxies, with merger properties similar to those that lead to LIRGs in the local Universe. 

\subsection{The physical mechanisms that trigger and quench a merger-induced starburst} \label{sec:discussion_sims}

As discussed in Section \ref{sec:intro}, quenching mechanisms driven by internal feedback, including AGN feedback and central starburst driven stellar feedback, are expected to lead to inside-out quenching. As shown in Fig. \ref{fig:radial_sfh}, the recent SFR decline in all three PSBs in our sample commenced first in the outer regions before propagating inwards. However, as discussed in previous sections, this is more to do with the early burst in the outer regions, likely driven by a past merger, than it is to do with the quenching mechanism. Our improved data and analysis allows us to dive into the case for spatially resolved quenching in more detail.  

Simulated merging galaxies often show multiple episodes of increased SFR that roughly coincide with or occur slightly after pericentre passages and the final black hole coalescence \citep[e.g.][]{johansson2009,perez2011,torrey2012,hopkins2013,zheng2020,moreno2021,petersson2023}. Here we compare and contrast two of these most recent simulations, which have a wide range of differences. It is beyond the scope of this discussion to dive into the details of the differences between the methods, but we aim to demonstrate how the combined power of simulations and observations can tell us about the physical processes of star formation and quenching likely dominating in the real Universe.  

\cite{zheng2020} (hereafter Z20) performed isolated merger simulations for a suite of galaxy progenitor initial conditions and merger orbitals including 1:1 and 1:3 mass ratios, in order to compare to the radial gradients in spectral indices found in post-starburst galaxies. They used the smoothed particle hydrodynamic (SPH) code GADGET-3 with a similar set up to the EAGLE cosmological simulation, stellar particles with mass resolution of $1.4\times10^5$M$_\odot$ and a gravitational smoothing length of 28\,pc for baryonic particles. The simulations included intense AGN feedback in order to evacuate the cold gas and result in post-starburst galaxies with weak $H\alpha$ emission lines, consistent with the PSBs studied in this paper. \cite{petersson2023} (hereafter P23) simulated two merging disc galaxies with a 1:2 mass ratio to reproduce the resolved star formation activity of shell galaxies, using the adaptive mesh refinement (AMR) code RAMSES and a maximum mass resolution of $2.4\times10^{4}$M$_\odot$ for stellar particles and a maximum physical resolution of 12\,pc in the most refined part of the mesh. These simulations did not include AGN feedback.

Z20 found that strong, global starbursts can be triggered following final coalescence, but only a minority of the initial conditions led to multiple bursts. They also split one of their simulations into concentric radial shells, showing that the major merger induced a single coeval starburst that was much stronger and slower quenched in the centre (their fig. 6). Thus, even with the presence of strong AGN feedback, their galaxies did not experience inside-out quenching. However, the simulations of Z20 are clearly not consistent with our observations, where a burst occurs earlier in the outer regions and propagates inwards. We find a burst decline time that is slower than in the inner regions, unlike in these simulations.

On the other hand, P23 found that the first starburst occurs shortly after the first pericentre passage in both the central and outer regions, with the SFR surface density in some outer regions exceeding pre-passage levels anywhere in the progenitors (their fig. 6). The starburst in the outer regions takes place in small fragmented pockets (their fig. 8), with the authors proposing tidal compression as the driving mechanism. Following final coalescence, a strong starburst occurs within the central $\sim1.5\;$kpc, while SFR in the outer regions falls gradually into quiescence. The tidal interactions lead to a morphological transformation of the system into a spheroid, and this transformation stabilises the gas and quenches the star formation following coalescence. With the absence of AGN feedback in these simulations, small amounts of molecular gas remain in the centre of the post-merger remnant, perhaps akin to that observed in many local PSBs \citep{french2015,french2023,alatalo2016,smercina2018,smercina2022,otter2022,rasmussen2026}. 

The stabilisation of gas against the gravitational collapse that form molecular clouds via dynamical suppression have been shown by simulations to be an effective way to quench star formation while retaining substantial gas reservoirs \citep{gensior2020,gensior2021,porter2026}. These studies predict that as a result of the suppression process, the molecular ISM are more smoothly distributed with a lack of locally dense clumps. This has been supported by observations of nearby early-type galaxies exhibiting smooth ISM morphologies, while the ISM in nearby spiral galaxies are more clumpy and asymmetrical \citep{davis2022}. Interestingly, despite their high molecular gas surface densities for their SFRs, some nearby PSBs appear to show low fractions of dense molecular gas relative to their total molecular gas \citep{french2023}, which could be a result of quenching via dynamical suppression.

Direct comparison between the P23 simulations and the observations presented here via forward modelling of the simulated spatially resolved optical spectra and molecular gas surface densities would be ideal to further identify similarities and differences, however even this simple comparison reveals that our SFH results better match the P23 simulations than those of Z20. 

Our results add to the growing evidence in the literature where mergers can effectively and rapidly shut down star formation in galaxies in the local Universe. However, detailed comparison with high resolution simulations suggest that this quenching occurs without the need of additional AGN feedback, in contrast to conclusions from lower resolution cosmological hydrodynamic simulations which conclude AGN feedback is significant in quenching PSBs \citep{quai2023,ludwig2025}. 

\subsection{Could AGNs have played a role?}

Some observational studies have found an elevated occurrence of AGNs in PSB galaxies than control samples \citep{yan2006,wild2007,yusef2014,krishna2025}. However, most of these AGNs appear to be low in luminosity \citep{lanz2022}. Conversely, other studies have found no evidence for this elevated AGN occurrence in PSBs \citep{depropris2014,meusinger2017,almaini2025}. This discrepancy could be due to differences in the selection of AGNs and PSBs \citep{krishna2025}, and may partially be due to a delay in AGN activity after a starburst found in some observations \citep{davies2007,schawinski2007,wild2010}. 

An elevated occurrence of AGNs has also been observed in post-merger galaxies \citep{ellison2019,ellison2025,li2023a}, but recent studies have found that among post-merger galaxies, those that show PSB signatures are similarly likely to host AGNs as non-PSBs \citep{ellison2022,li2023}. This suggests that while the increased gas fuelling during mergers may have promoted the ``switching on'' of AGNs, the AGNs have limited impact in directly rapidly quenching the merger-induced starbursts.

While AGN feedback, if effective, is often expected to act inside-out \citep[e.g.][]{zubovas2016,bluck2020}, \cite{davies2022} used simulations to demonstrate that outflows from an AGN could expel a large fraction of the galaxy's circumgalactic medium, thus halting the galaxy's supply of inflowing gas, leading to outside-in quenching. However, this process quenches over several Gyrs, which is much too slow to create the PSB signatures seen in the typical local PSB galaxy.

Nonetheless, we search for signs of the presence of AGNs in our three galaxies.
From the MaNGA data, no broad lines can be observed in the central spaxels of the three PSBs studied. We construct BPT and WHAN diagrams for all spaxels in the three galaxies \citep{baldwin1981,cid_fernandes2011}, finding no spaxels in any galaxies to lie in the Seyfert regions. Only a portion of the central regions of 12514-3702 are placed in the ``weak AGN'' portion of the WHAN diagram. The MaNGA AGN catalogue from \cite{alban2023} classified 7976-1902 and 12514-3702 as LIERs/LINERs, but a considerable number of spaxels in the outskirts in both galaxies lie on similar loci in the BPT and WHAN diagrams as their central spaxels. Therefore, the LIER/LINER signatures are not restricted to the centre, indicating their origin are likely not from AGNs. Additionally, we cross matched the galaxies with the XMM Slew \citep{salvato2018} and eROSITA \citep[see section 7.2 of][]{SDSS_DR19} value added catalogues of X-ray sources. Two of the galaxies (7965-1902 and 12067-3701) are within the footprint of the XMM Slew catalogue, while the third (12514-3702) is covered by eROSITA. We find no matching X-ray sources within one arcsecond of any of the galaxies, indicating no X-ray AGN presence. In summary, all three galaxies are unlikely to host AGNs. 

\section{Conclusions}
To investigate the origin of post-starburst galaxies, we fit the spatially resolved spectra of three local face-on PSBs with extended central PSB regions from MaNGA using a Bayesian hierarchical model. Within this model, we assumed functional forms to describe the radial profiles of local stellar mass surface density, burst age, pre-burst stellar metallicity, extinction in $V$-band and velocity dispersion. The usage of the hierarchical model to fit all spatially resolved spectra under a joint framework has allowed for population properties to be simultaneously constrained with local properties, and improved the estimation precision on the latter.

We produced radial profiles and maps of 10 key galaxy properties, which led to six main conclusions:
\begin{enumerate}
    \item All three galaxies show a distinct two-phase outside-in sequence of starburst and quenching, where the outer regions first experienced a weaker starburst that quenched slower before the central regions experienced a stronger, more rapidly quenching starburst.
    \item The earlier, outer starburst induced a weak increase in stellar metallicity, while the later, central starburst induced a strong increase in stellar metallicity.
    \item The outside-in starburst sequence is consistent with a recent merger, where the first pericentre passage triggered starbursts predominantly in the outer regions of the galaxies, but the later coalescence triggers predominantly a central starburst.
    \item We observe non-axisymmetric features in the maps of burst mass fraction and dust attenuation in all galaxies, which we consider as further evidence of tidal effects during the recent merger.
    \item Comparison to observations of SFR enhancement in ongoing mergers suggests that the PSBs progenitors are similar to local LIRGs, i.e. merging galaxies with high gas fractions. 
    \item Our results are more consistent with binary merger simulations where rapid quenching is caused by the consumption of dense ISM gas through star formation, followed by stabilisation of the remaining gas due to the remnants spheroidal morphology (morphological quenching), rather than the ejection of star-forming gas through extensive AGN feedback.
\end{enumerate}

The measured resolved properties and SFHs hold a wealth of information that we have yet to fully explore. For example, the side-to-side gradients in quenching timescale ($\tau_{1/2}$) in the central recent starbursts of 7965-1902 and 12514-3702 could be related to directional feedback or large-scale shocks. The BHM application in this study assumes the same SFH model throughout galaxies, which restricts its application to galaxies that are likely to have PSB SFHs in most/all regions. To broaden the application of the BHM method, future work could explore less restrictive SFH models (e.g. non-parametric models) or allow the switching of SFH forms through mixture models. Lastly, direct comparison between observations and forward-modelled hydrodynamical simulations with high spatial resolution, both concerning galaxy mergers and other mechanisms that might lead to starbursts and rapid quenching, are needed to properly compare predicted resolved SFHs to observations. This will better constrain the physical processes that led to these galaxies' recent evolution.

\section*{Acknowledgements}

HL thanks Johannes Lange for help with using \textsc{Nautilus}, Ian Czekala for help with BHM theory, and Am\'elie Santonge and Andrew Cameron for useful discussions and comments. VW thanks James Nightingale for useful discussions around implementing BHM for spectral fitting of galaxies. 
HL and ACC acknowledge support from a UKRI Frontier Research Grantee Grant (PI Carnall; grant reference EP/Y037065/1).
VW acknowledges Science and Technologies Facilities Council (STFC) grants ST/V000861/1 and ST/Y00275X/1, and Leverhulme Research Fellowship RF-2024-589/4. ALR acknowledges support from Leverhulme Early Career Fellowship ECF-2024-186. PHJ acknowledges the support by the European Research Council via ERC Consolidator grant KETJU (no. 818930).

Funding for the Sloan Digital Sky 
Survey IV has been provided by the 
Alfred P. Sloan Foundation, the U.S. 
Department of Energy Office of 
Science, and the Participating 
Institutions. 

SDSS-IV acknowledges support and 
resources from the Center for High 
Performance Computing  at the 
University of Utah. The SDSS 
website is www.sdss4.org.

SDSS-IV is managed by the 
Astrophysical Research Consortium 
for the Participating Institutions 
of the SDSS Collaboration including 
the Brazilian Participation Group, 
the Carnegie Institution for Science, 
Carnegie Mellon University, Center for 
Astrophysics | Harvard \& 
Smithsonian, the Chilean Participation 
Group, the French Participation Group, 
Instituto de Astrof\'isica de 
Canarias, The Johns Hopkins 
University, Kavli Institute for the 
Physics and Mathematics of the 
Universe (IPMU) / University of 
Tokyo, the Korean Participation Group, 
Lawrence Berkeley National Laboratory, 
Leibniz Institut f\"ur Astrophysik 
Potsdam (AIP),  Max-Planck-Institut 
f\"ur Astronomie (MPIA Heidelberg), 
Max-Planck-Institut f\"ur 
Astrophysik (MPA Garching), 
Max-Planck-Institut f\"ur 
Extraterrestrische Physik (MPE), 
National Astronomical Observatories of 
China, New Mexico State University, 
New York University, University of 
Notre Dame, Observat\'ario 
Nacional / MCTI, The Ohio State 
University, Pennsylvania State 
University, Shanghai 
Astronomical Observatory, United 
Kingdom Participation Group, 
Universidad Nacional Aut\'onoma 
de M\'exico, University of Arizona, 
University of Colorado Boulder, 
University of Oxford, University of 
Portsmouth, University of Utah, 
University of Virginia, University 
of Washington, University of 
Wisconsin, Vanderbilt University, 
and Yale University.

\textit{Software:} \textsc{Astropy} \citep{astropy}, \textsc{Bagpipes} \citep{bagpipes2018,bagpipes2019}, \textsc{Celerite2} \citep{celerite,celerite2}, \textsc{Daft} \citep{daft}, \textsc{Marvin} \citep{marvin}, \textsc{Matplotlib} \citep{matplotlib}, \textsc{Nautilus} \citep{nautilus}, \textsc{Numba} \citep{numba}, \textsc{Numpy} \citep{numpy}, \textsc{pipes\_vis} \citep{pipes_vis}, \textsc{Scipy} \citep{scipy}, \textsc{Seaborn} \citep{seaborn}

For the purpose of open access, the author has applied a Creative Commons Attribution (CC BY) licence to any Author Accepted Manuscript version arising.

\section*{Data Availability}
All utilised MaNGA data are publicly available at the SDSS database \url{https://www.sdss4.org/dr17/} or through \texttt{Marvin} at \url{https://dr17.sdss.org/marvin/}. Maps of estimated properties shown in Figs. \ref{fig:results_7965} to \ref{fig:results_12514} along with the Voronoi binned data are available at \url{https://doi.org/10.17630/009bb751-d000-4404-b397-9e4ed6dd72bf}. Python scripts to recreate the figures in Section \ref{sec:results} are available at \url{https://github.com/HinLeung622/MaNGA_PSB_resolved_SFH}.



\bibliographystyle{mnras}
\bibliography{biblist} 




\appendix

\section{Detailed derivation of Bayesian hierarchical modelling theory} \label{apx:derivations1}
Here, we provide a brief derivation of BHM up to Equation \ref{eq:Hbayes_main}. For more detailed derivations, the reader is referred to \cite{bernardo2000_book} and \cite{gelman2004_bayesbook}, and more recently \cite{andreon2015_book} and \cite{thrane2019}.

As written in Section \ref{sec:theory}, the joint posterior distribution of $\vec{\alpha}$ and $\mathbf{\Theta}$ (all $\vec{\theta_i}$) is given by
\begin{equation}
    P(\vec{\alpha},\mathbf{\Theta}|\mathbf{D}) = \frac{P(\mathbf{D}|\mathbf{\Theta},\vec{\alpha}) \; P(\mathbf{\Theta},\vec{\alpha})}{P(\mathbf{D})} \; .
\end{equation}

Using the definition of conditional probability $P(A|B) = P(A, B)/P(B)$, it follows that
\begin{equation}
    P(\vec{\alpha},\mathbf{\Theta}|\mathbf{D}) = \frac{P(\mathbf{D}|\mathbf{\Theta},\vec{\alpha}) \; P(\mathbf{\Theta}|\vec{\alpha}) \; P(\vec{\alpha})}{P(\mathbf{D})} \; .
\end{equation}
$P(\vec{\alpha})$ is the prior on the hyper-parameters, known as the hyper-prior. $P(\mathbf{\Theta}|\vec{\alpha})$ is the conditional prior for the parameters $\mathbf{\Theta}$. $P(\mathbf{D}|\mathbf{\Theta},\vec{\alpha})$ is the likelihood of the observed data for all $N_V$ units given $\mathbf{\Theta}$, which has no direct dependence on $\vec{\alpha}$; the hyper-parameters $\vec{\alpha}$ only affect $\mathbf{D}$ through $\mathbf{\Theta}$. 

By asserting that the observations $\mathbf{D}$ are independent given $\mathbf{\Theta}$ and $\vec{\alpha}$, the joint likelihood can be rewritten as the product of each region's likelihood:
\begin{equation}
    P(\mathbf{D}|\mathbf{\Theta}) = \prod^{N_V}_{i=1} P(\vec{D_i}|\vec{\theta_i}) \; .
\end{equation}
$P(\mathbf{\Theta}|\vec{\alpha})$ is the joint conditional prior representing the probability density function (PDF) of $\mathbf{\Theta}$ given the hyper-priors $P(\vec{\alpha})$. Due to no direct cross dependencies between the $\vec{\theta_i}$, for $i=1, \dots, N_V$, it can be rewritten as
\begin{equation}\label{eq:Hbayes_theory4}
    P(\mathbf{\Theta}|\vec{\alpha}) = \prod^{N_V}_{i=1} P(\vec{\theta_i}|\vec{\alpha}) \; .
\end{equation}
Therefore, we can write the joint posterior distribution of $\mathbf{\Theta}$ and $\vec{\alpha}$ given data $\mathbf{D}$ as
\begin{equation} \label{eq:Hbayes_main2}
    P(\vec{\alpha},\mathbf{\Theta}|\mathbf{D}) = \frac{P(\vec{\alpha}) \prod^{N_V}_{i=1}\Big[P(\vec{D_i}|\vec{\theta_i}) \; P(\vec{\theta_i}|\vec{\alpha})\Big]}{P(\mathbf{D})} \; ,
\end{equation}
which matches Equation \ref{eq:Hbayes_main}.

For examples applications of BHM in astronomy, the reader can refer to the growing literature in many fields, including in cosmology \citep[e.g.][]{alsing2017,hinton2019}, gravitational waves \citep[e.g.][]{adams2012,thrane2019}, stellar observation and evolution \citep[e.g.][]{leistedt2017,hall2019,lyttle2021}, and exoplanets \citep[e.g.][]{hogg2010,forman-mackey2014,czekala2019}. In extragalactic astronomy, BHM has been used to improve the precision of photometric redshift estimates \citep{leistedt2016,leistedt2023,sanchez2019,rau2020,alarcon2020}, to map the dust properties of nearby galaxies through SED fitting \citep{galliano2018,lamperti2019,chang2021}, to improve constraints on the star-forming main sequence \citep{curtis-lake2021,sandles2022} and to study the stellar-halo mass relation with galaxy-galaxy lensing data \citep{sonnenfeld2018}. Many studies have commented on the advantages of BHM in degeneracy-breaking \citep{juvela2013,galliano2018}, uncertainty propagation \citep{curtis-lake2021,sandles2022}, and reduction of bias in the results \citep{sonnenfeld2018}.

Leveraging the idea that physical processes in galaxies span a broad range of physical scales, BHM have also seen recent applications in measuring resolved galaxy properties, such as mapping stellar ages from photometry \citep{sanchez-gil2019}. \cite{gonzalez-gaitan2019} proposed a hierarchical framework that treats galaxy properties (age, mass, metallicity, dust extinction) as unobserved Gaussian Markov random fields, and measured resolved maps of these properties from IFS surveys CALIFA and PISCO. \cite{ditrani2024} modelled VLT/MUSE IFS observations of the Cartwheel galaxy with a two-stage hierarchical approach, where the posterior distributions of local properties measured from sub-regions are combined to yield the posterior distributions of the region's global property. These studies, in particular \cite{gonzalez-gaitan2019}, demonstrate the advantages of BHM, such as increased spatial resolution when given poor SNR data, and combining multiple pointings and observations of different spatial resolutions in a formal model.

\section{Detailed derivations of the rejection sampling step} \label{apx:derivations2}

Rejection sampling is often used instead of direct sampling from a target PDF $f(\theta)$ when direct sampling is non-trivial or computationally expensive. The method relies on the usage of $g(\theta)$, which is a PDF we can easily draw samples from directly. Then, the samples drawn can be accepted or rejected using the acceptance probability $f(\theta)/g(\theta)$ to indirectly sample from the PDF $f(\theta)$. Typically, the target PDF $f(\theta)$ is some posterior distribution of interest ($P(\theta|\vec{D})$), and $g(\theta)$ is a general curve that \textit{envelopes} the posterior distribution \citep{stanford1994,gelman2004_bayesbook}.

For the $i$-th Voronoi bin and the $k$-th posterior sample of hyper-parameters from \textit{stage 2}, the target PDF $f(\vec{\theta_i})$ is
\begin{align}
    f(\vec{\theta_i}) = P(\vec{\theta_i}|\vec{D_i},\vec{\alpha_k}) &= \frac{P(\vec{D_i}|\vec{\theta_i},\vec{\alpha_k}) \; P(\vec{\theta_i}|\vec{\alpha_k})}{P(\vec{D_i|\vec{\alpha_k}})} \label{eq:rejection_samp_step1}\\
     &= \frac{P(\vec{D_i}|\vec{\theta_i}) \; P(\vec{\theta_i}|\vec{\alpha_k})}{P(\vec{D_i}|\vec{\alpha_k})} \; , \label{eq:rejection_samp_step2}
\end{align}
where the $\vec{\alpha_k}$ dependence of the likelihood $P(\vec{D_i}|\vec{\theta_i},\alpha_k)$ in Equation \ref{eq:rejection_samp_step1} is removed to give Equation \ref{eq:rejection_samp_step2}, similar to that in Equation \ref{eq:theory_1}.

The ``simple'' PDF $g(\vec{\theta_i})$ from which we sample is the posterior distribution from \textit{stage 1}, which is given in Equation \ref{eq:theory_non_hierarchical}. If we take the ratio $f(\vec{\theta_i})/g(\vec{\theta_i})$ to obtain the acceptance probability, and use the property that the likelihood from the hierarchical model is equal to the likelihood from the non-hierarchical model ($P(\vec{D_i}|\vec{\theta_i}) = P(\vec{D_i}|\vec{\theta_i^\mathrm{NP}})$), we have:
\begin{align}
    \frac{f(\vec{\theta_i})}{g(\vec{\theta_i})} &= \frac{P(\vec{D_i}|\vec{\theta_i}) \; P(\vec{\theta_i}|\vec{\alpha_k})}{P(\vec{D_i|\vec{\alpha_k}})} \cdot \frac{P(\vec{D_i})}{P(\vec{D_i}|\vec{\theta_i^\mathrm{NP}}) \; P(\vec{\theta_i^\mathrm{NP}})} \\
    &= \frac{P(\vec{\theta_i}|\vec{\alpha_k})}{P(\vec{\theta_i^\mathrm{NP}})} \; \frac{P(\vec{D_i})}{P(\vec{D_i}|\vec{\alpha_k})} \; .
\end{align}

The terms $P(\vec{D_i})$ and $P(\vec{D_i}|\vec{\alpha_j})$ are scalars that adjust the acceptance probability $P(\vec{\theta_i}\;\mathrm{is\;accepted}) = f(\vec{\theta_i})/g(\vec{\theta_i})$, and do not depend on $\theta_i$. To maximize sampling efficiency, we maximize the overall probability of acceptance for all $\vec{\theta_i}$. This can be achieved by lowering the the ``simple'' distribution $g(\vec{\theta_i})$. However, $g(\vec{\theta_i})$ can not be lowered beyond the point where it no longer ``envelops'' the target distribution $f(\vec{\theta_i})$, i.e. when $P(\vec{\theta_i}\;\mathrm{is\;accepted}) > 1$ for any $\vec{\theta_i}$. Thus, $g(\vec{\theta_i})$ can only be lowered until the maximum probability of acceptance is equal to 1. The acceptance probability of our rejection sampler is given by the fraction $f(\vec{\theta_i})/g(\vec{\theta_i})$ divided by the supremum of the fraction, which has the form
\begin{align}
    P(\vec{\theta_i}\;\mathrm{is\;accepted}) &= \frac{\frac{P(\vec{\theta_i}|\vec{\alpha_k})}{P(\vec{\theta_i^\mathrm{NP}})} \; \frac{P(\vec{D_i})}{P(\vec{D_i}|\vec{\alpha_k})}}{\sup_{\vec{\theta_i}} \Bigg( \frac{P(\vec{\theta_i}|\vec{\alpha_k})}{P(\vec{\theta_i^\mathrm{NP}})} \; \frac{P(\vec{D_i})}{P(\vec{D_i}|\vec{\alpha_k})} \Bigg)} \\
    &= \frac{P(\vec{\theta_i}|\vec{\alpha_k})/P(\vec{\theta_i^\mathrm{NP}})}{\sup_{\vec{\theta_i}} \Big[ P(\vec{\theta_i}|\vec{\alpha_k}) / P(\vec{\theta_i^\mathrm{NP}}) \Big] } \; .
\end{align}
If we replace $\vec{\theta_i}$ with the parameters that directly depend on the hyper-parameters ($\vec{\phi_i}$), we have Equation \ref{eq:acceptance_prob}. Given that the sampler from \textit{stage 2} is converged and can be treated as a collapsed Gibbs sampler \citep{van_dyk2008}, it is reasonable to assume that the samples drawn through this rejection sampling method using the acceptance probability above are the same as samples drawn directly from $P(\vec{\theta_i}|\vec{D_i},\vec{\alpha_k})$.

\section{Uncertainty maps} \label{apx:uncertainty_maps}
\begin{figure*}
    \centering
    \includegraphics[width=\textwidth]{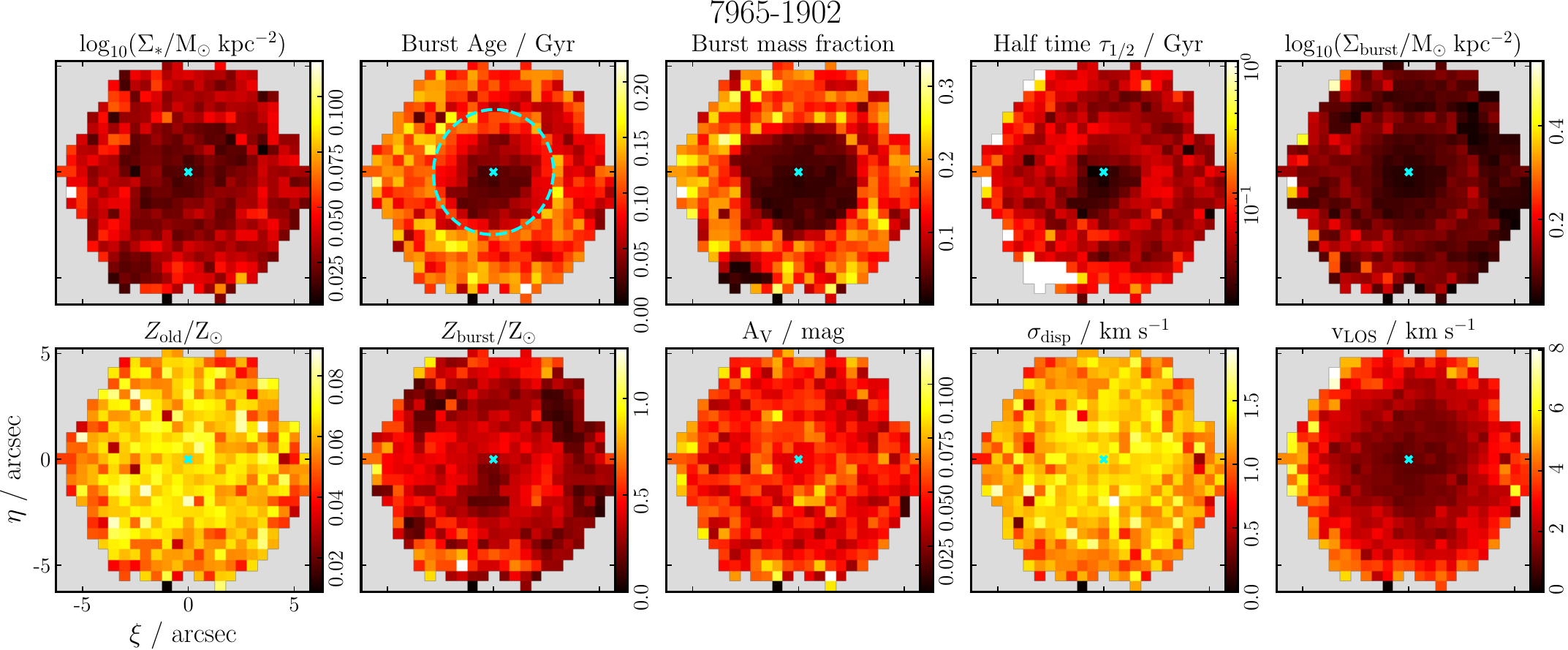}
    \includegraphics[width=\textwidth]{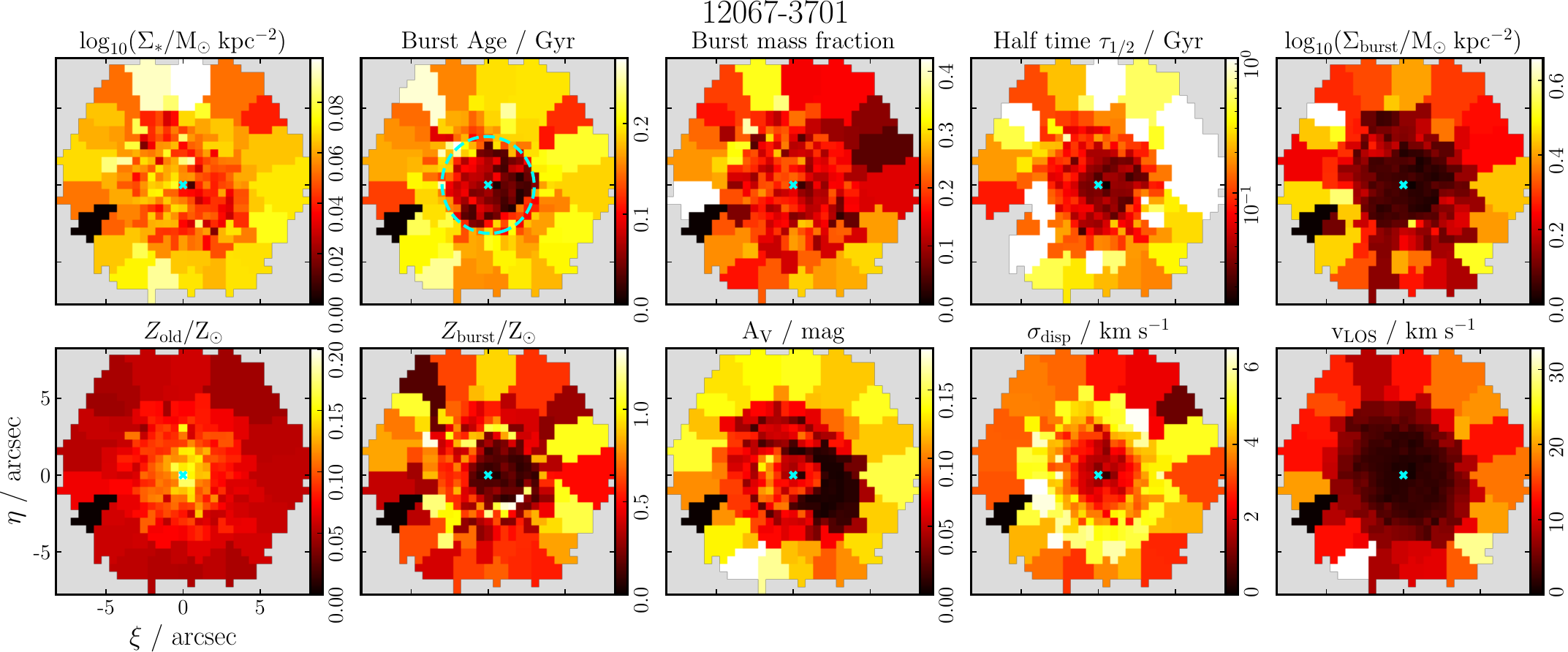}
    \includegraphics[width=\textwidth]{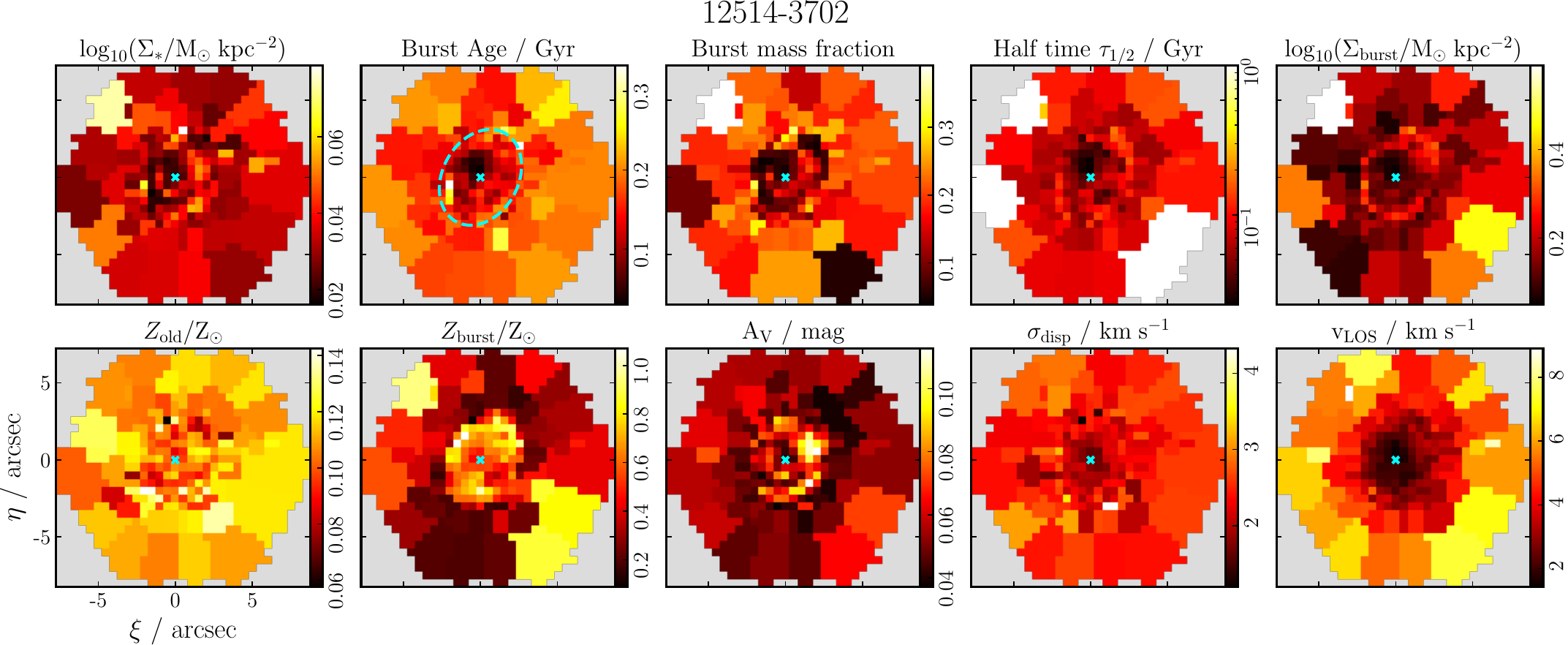}
    \caption{Maps of $1\sigma$ uncertainties of the resolved properties in Figs. \ref{fig:results_7965}, \ref{fig:results_12067} and \ref{fig:results_12514}.}
    \label{fig:err_maps}
\end{figure*}

For completeness, in Fig. \ref{fig:err_maps} we include the $1\sigma$ uncertainties maps of the resolved properties shown in Figs. \ref{fig:results_7965}, \ref{fig:results_12067} and \ref{fig:results_12514}.

\bsp	
\label{lastpage}
\end{document}